# Evolving ribonucleocapsid assembly/packaging signals in the genomes of the human and animal coronaviruses: targeting, transmission and evolution

Vladimir R. Chechetkin[a]* and Vasily V. Lobzin[b]

[a]*Engelhardt Institute of Molecular Biology of Russian Academy of Sciences, Vavilov str., 32, Moscow, Russia*

[b]*School of Physics, University of Sydney, Sydney, NSW 2006, Australia*

______________________________

*Corresponding author. E-mail addresses:* chechet@eimb.ru; vladimir_chechet@mail.ru

Tel.: +7 499 135 9895. Fax: +7 499 135 1405 (V.R. Chechetkin).

## Abstract

A world-wide COVID-19 pandemic intensified strongly the studies of molecular mechanisms related to the coronaviruses. The origin of coronaviruses and the risks of human-to-human, animal-to-human, and human-to-animal transmission of coronaviral infections can be understood only on a broader evolutionary level by detailed comparative studies. In this paper, we studied ribonucleocapsid assembly-packaging signals (RNAPS) in the genomes of all seven known pathogenic human coronaviruses, SARS-CoV, SARS-CoV-2, MERS-CoV, HCoV-OC43, HCoV-HKU1, HCoV-229E, and HCoV-NL63 and compared them with RNAPS in the genomes of the related animal coronaviruses including SARS-Bat-CoV, MERS-Camel-CoV, MHV, Bat-CoV MOP1, TGEV, and one of camel alphacoronaviruses. RNAPS in the genomes of coronaviruses were evolved due to weakly specific interactions between genomic RNA and N proteins in helical nucleocapsids. Combining transitional genome mapping and Jaccard correlation coefficients allows us to perform the analysis directly in terms of underlying motifs distributed over the genome. In all coronaviruses RNAPS were distributed quasi-periodically over the genome with the period about 54 nt biased to 57 nt and to 51 nt for the genomes longer and shorter than that of SARS-CoV, respectively. The comparison with the experimentally verified packaging signals for MERS-CoV, MHV, and TGEV proved that the distribution of particular motifs is strongly correlated with the packaging signals. We also found that many motifs were highly conserved in both characters and positioning on the genomes throughout the lineages that make them promising therapeutic targets. The mechanisms of encapsidation can affect the recombination and co-infection as well.

## Keywords

## List of Abbreviations

Bat-CoV MOP1, miniopterus bat coronavirus 1, HCoV-229E, human coronavirus 229E; HCoV-HKU1, human coronavirus HKU1; HCoV-NL63, human coronavirus NL63; HCoV-OC43, human coronavirus OC43; JC, Jaccard coefficient; MERS-CoV, Middle East respiratory syndrome coronavirus; MHV, murine hepatitis virus; N proteins, nucleocapsid proteins; nt, nucleotides; RNAPS, ribonucleocapsid assembly/packaging signals; SARS-CoV, severe acute respiratory syndrome coronavirus; SARS-CoV-2, severe acute respiratory syndrome coronavirus 2; sgmRNA, subgenomic messenger RNA; TAMGI, transitional automorphic mapping of the genome on itself; TGEV, transmissible gastroenteritis virus; UTR, untranslated region

# 1. Introduction

The COVID-19 pandemic inspired worldwide crisis by the rapid spread of infectious disease caused by severe acute respiratory syndrome coronavirus 2 (SARS-CoV-2). There are strong indications that coronaviral infections will become a permanent significant factor affecting the life of human society. Currently, coronaviruses cause 30% of upper and lower respiratory tract infections in humans, SARS-CoV, SARS-CoV-2 and MERS-CoV being the cause of heavy diseases and deaths. Like the other diseases caused by coronaviruses, COVID-19 is thought to be of zoonotic origin (Contini et al., 2020; Decaro & Lorusso, 2020; Mahdy et al, 2020; Swelum et al., 2020; Zhou et al., 2020). The assessment of risks of human-to-human, animal-to-human, and human-to-animal transmission of coronaviral infections and development of efficient medications and vaccines against the coronaviruses need the knowledge of main molecular mechanisms in the virus life cycle and virus-host interactions (Mishra & Tripathi, 2021; O'Leary et al., 2020; Rabi et al., 2020; Saxena, 2020).

The long (about 30,000 nt) non-segmented plus-sense single-stranded RNA genome of the coronaviruses is packaged within a filament-like helical nucleocapsid, while the whole ribonucleocapsid is packaged within a membrane envelope with spike glycoproteins (Chang et al., 2014; Chen et al., 2007; Gui et al., 2017; Masters, 2019; Neuman & Buchmeier, 2016). The nucleocapsid (N) proteins of coronaviruses provide one of the promising therapeutic targets as they are critical for viral replication and assembly and can be attributed to the most conserved proteins (Bai et al., 2021; Chang et al., 2014; 2016; Dinesh et al., 2020; Dutta et al., 2020; Gao et al., 2021; Kwarteng et al., 2020; Lin et al., 2020; Matsuo, 2021; Peng et al., 2020; Tilocca et al., 2020; Yadav et al., 2020; Ye et al., 2020; Zinzula et al., 2021). The cryogenic electron microscopy (cryo-EM) has revealed that the ribonucleocapsid of SARS-CoV is helical with an outer diameter of 16 nm and an inner diameter of 4 nm (Chang et al., 2014). The turn of the nucleocapsid is composed of two octamers polymerized from dimeric N proteins (Chen et al., 2007). The pitch for the SARS-CoV nucleocapsid is 14 nm. The packaging of the SARS-CoV ssRNA genome near internal surface of helical nucleocapsid with such parameters should correspond to 54–56 nt per helical turn (or 6.75–7 nt per N protein) (Chang et al., 2014; Chechetkin & Lobzin, 2020). The parameters of the ribonucleocapsid for murine hepatitis virus (MHV) are close to those of SARS-CoV (Gui et al., 2017) in accordance with evolutionary conservation of the main structural characteristics of N proteins. Due to the transitional symmetry of a helix, the weakly specific cooperative interaction between ssRNA and nucleocapsid proteins should lead to the natural selection of specific quasi-periodic motifs in the related genomic sequences. Indeed, the quasi-periodic motifs with the period close to 54 nt were detected and appeared to be strongly pronounced in the genomes of SARS-CoV and SARS-CoV-2 (Chechetkin & Lobzin, 2020). The ribonucleocapsid assembly/packaging signals (RNAPS) in RNA genomes together with variations in N protein conformations ensure the specificity of encapsidation. In this paper we studied RNAPS in the genomes of the known endemic and pandemic human coronaviruses and compared them with RNAPS for a batch of related genomes for animal coronaviruses. The analysis of RNAPS is performed directly in terms of underlying motifs distributed over the genome and in the context of the general genome organization. Such RNAPS may serve for therapeutic targeting or for the assessment of evolutionary divergence between the viruses. They may also shed additional light on the mechanisms of recombination and co-infection for human coronaviruses as well as on the basic problems of viral molecular evolution.

## 2. Theory and methods

### 2.1. Correlational motifs and transitional genome mapping

In our study, we chose the scheme of analysis providing the output data directly in terms of related motifs distributed over the genome which are most important for the genetic and medical applications. The primary objects in our approach are correlational motifs. Generally, they cannot be reduced to the sparse or tandem repeats with gaps and alignments. As periodic features produce decaying or persistent oscillations in correlational motifs separated by distances multiple to the period, the periodic features can also be detected by the suggested method. The method can detect and reconstruct tandem repeats (both complete and incomplete) as well. The distribution of correlational motifs over the genome provides information about large-scale genome organization. As the correlational motifs are approximately robust with respect to point mutations and indels, they are especially suitable for the study of viral genomes with inherent high frequency of mutations. The characteristic correlational and quasi-repeating motifs can be used, e.g., for therapeutic targeting, for subtyping of viruses or for the assessment of evolutionary divergence between species.

The algorithm for transitional automorphic mapping of the genome on itself (TAMGI) (Chechetkin & Lobzin, 2020, 2021) used throughout this paper is defined as follows. Let a nucleotide $N_{m,\alpha}$ of the type α be positioned at a site $m$ of the genomic sequence. Then, a pair of $s$-neighbors, $N_{m-s}$ and $N_{m+s}$, is searched for around $N_{m,\alpha}$. The nucleotide $N_{m,\alpha}$ will be retained if it has at least one $s$-neighbor $N_{m-s,\alpha}$ or $N_{m+s,\alpha}$ of the same type and be replaced by void otherwise (denoted traditionally by hyphen). All $s$-neighbors of the same type, $N_{m-s,\alpha}$ or/and $N_{m+s,\alpha}$, should also be retained. The separation distance $s$ is called the step of TAMGI. The resulting sequence after TAMGI with the step $s$ is composed of the nucleotides of four types (A, C, G, T) and the hyphens "-" denoting voids. We will call such sequences TAMGI components and denote as $\Gamma_s$, whereas the whole genomic sequence will be denoted as $\Gamma$. The TAMGI components can be studied separately or be united, $\Gamma_{s_1} \bigcup \Gamma_{s_2}$, or intersected, $\Gamma_{s_1} \bigcap \Gamma_{s_2}$.

Further analysis after TAMGI is reduced to the study of all complete words of length $k$ ($k$-mers) composed only of nucleotides (voids within the complete words are prohibited) and surrounded by the voids "-" at 5'- and 3'-ends, $-N_k-$. By definition, the complete words are non-overlapping. At the next stage, the mismatches with hyphens to the complete words can be studied. The correlation motifs (or TAMGI motifs) are defined as a set of $k$-mers generated by TAMGI.

To avoid end effects and to ensure homogeneity of the mapping, the linear genomes will always be circularized,

$$N^c_{m,\alpha} = \begin{cases} N_{m,\alpha}, & \text{if } 1 \leq m \leq M; \\ N_{m-M,\alpha}, & \text{if } M+1 \leq m \leq 2M-1, \end{cases} \quad (1)$$

where $N_{m,\alpha}$ denotes the nucleotide of the type $\alpha \in$ (A, C, G, T) occupying the site $m$ and $M$ is the genome length. The circularized version of TAMGI can also be described as follows. (i) Circularize a linear genome and superimpose two identical circular genomes over each other. (ii) Rotate clockwise one of the genomes on a step $s$ and count all coincidences between the two genomes. (iii) Rotate counterclockwise one of the genomes on a step $s$ and count all coincidences between the two genomes. (iv) Unite all coincidences into one sequence and fill the voids by hyphens.

The theory and simulations show that for even $M$ the step $M/2$ should be considered apart from the other steps. Therefore, the range of steps can be chosen from 1 to

$$L = \begin{cases} [M/2] \text{ for odd } M; \\ M/2 - 1 \text{ for even } M, \end{cases} \quad (2)$$

where the brackets denote the integer part of the quotient. Any sequence can be expanded via the complete set of TAMGI components with the steps from $s = 1$ to $[M/2]$,

$$\Gamma = \bigcup_{s=1}^{[M/2]} \Gamma_s . \quad (3)$$

This means that $\Gamma_s$ can be considered as the generalized genome coordinates, whereas the highest $\Gamma_s$ can be associated with the principal components related to the genome organization.

The frequencies of nucleotides after TAMGI with the step $s$,

$$\varphi_{\alpha,s} = N_{\alpha,s}/M; \quad \varphi_{total,s} = \sum_{\alpha} \varphi_{\alpha,s}, \quad (4)$$

should be properly normalized to assess their statistical significance. The normalization ought to be performed against the counterpart characteristics in the random sequences of the same nucleotide composition. The partial fractions of nucleotides after TAMGI for the randomly reshuffled genomic sequences are given by

$$\Phi_\alpha = \varphi_\alpha^2 (2 - \varphi_\alpha); \quad \Phi_{total} = \sum_{\alpha} \Phi_\alpha, \quad (5)$$

where $\varphi_\alpha$ is the frequency of nucleotides of the type $\alpha$ retained under reshuffling (the detailed theory for TAMGI, including the derivation of Eq. (5), can be found in the paper by Chechetkin & Lobzin (2021)).

The frequencies given by Eq. (5) are independent of steps $s$. For random sequences, the variances for the frequencies defined by Eq. (5) obey the binomial distribution,

$$\sigma^2(\Phi_\alpha) = \Phi_\alpha(1-\Phi_\alpha)/M; \quad \sigma^2_{total} = \sum_\alpha \sigma^2(\Phi_\alpha). \tag{6}$$

The total frequency defined by Eq. (4) can be presented in terms of a normalized deviation,

$$\kappa_s = (\varphi_{total,s} - \Phi_{total})/\sigma_{total}. \tag{7}$$

In random sequences, the deviations defined by Eq. (7) are governed by the Gaussian statistics and the deviations for the different steps $s$ are approximately independent. The steps associated with the pronounced deviations of high statistical significance are of primary interest. For quasi-periodic motifs with a period $p$, the deviations defined by Eq. (7) should reveal approximately equidistant peaks at the steps $s = p, 2p, \ldots$ A part of statistically significant deviations may be associated with correlations between nucleotides.

## *2.2. Correlations between TAMGI components and Jaccard coefficient*

The mutual selection of functional motifs in viral genomes, their merging and decaying during molecular evolution strongly affect the molecular mechanisms of the virus life cycle. To assess such effects, we will use the Jaccard correlation coefficient (Baharav et al., 2020; Chung et al., 2019; Jaccard, 1912; Vorontsov et al., 2013). The number of nucleotides in a sequence obtained after TAMGI with the step $s$ will be denoted as

$$|\Gamma_s| = N_s, \tag{8}$$

whereas the total related frequencies are determined by Eq. (4). The resulting numbers of nucleotides for the union and intersection of TAMGI components are denoted as

$$|\Gamma_{s_1} \bigcup \Gamma_{s_2}| \equiv |\Gamma_{s_1 \bigcup s_2}| = N_{s_1 \bigcup s_2}, \tag{9}$$

$$|\Gamma_{s_1} \bigcap \Gamma_{s_2}| \equiv |\Gamma_{s_1 \bigcap s_2}| = N_{s_1 \bigcap s_2}, \tag{10}$$

and are related by the equality

$$|\Gamma_{s_1} \bigcup \Gamma_{s_2}| = |\Gamma_{s_1}| + |\Gamma_{s_2}| - |\Gamma_{s_1} \bigcap \Gamma_{s_2}|. \tag{11}$$

The corresponding frequencies are calculated by the standard definitions,

$$\varphi_{total; s_1 \bigcup s_2} = N_{s_1 \bigcup s_2}/M = \sum_\alpha \varphi_{\alpha; s_1 \bigcup s_2}, \tag{12}$$

$$\varphi_{total; s_1 \bigcap s_2} = N_{s_1 \bigcap s_2}/M = \sum_\alpha \varphi_{\alpha; s_1 \bigcap s_2}. \tag{13}$$

Then, Jaccard correlation coefficient is defined as

$$J_{s_1 \wedge s_2} = \varphi_{total; s_1 \cap s_2} / \varphi_{total; s_1 \cup s_2} . \qquad (14)$$

The related frequencies for random sequences are given by

$$\Phi_{total, s_1 \cup s_2} = 2\Phi_{total} - \Phi_{total, s_1 \cap s_2}; \quad \Phi_{total, s_1 \cap s_2} = \sum_\alpha \Phi_\alpha^2 / \varphi_\alpha \qquad (15)$$

(see for notations Eq. (5)). This yields the following approximate expression for the Jaccard coefficient,

$$<J_{s_1 \wedge s_2}>_{random} = \left(\sum_\alpha \Phi_\alpha^2 / \varphi_\alpha\right) / \left(2\Phi_{total} - \sum_\alpha \Phi_\alpha^2 / \varphi_\alpha\right), \qquad (16)$$

whereas the corresponding variance can be assessed as

$$\sigma^2(J_{s_1 \wedge s_2}) = \sum_\alpha \sigma_\alpha^2(J_{s_1 \wedge s_2});$$
$$\sigma_\alpha^2(J_{s_1 \wedge s_2}) \approx 4\sigma^2(\Phi_\alpha)(2\Phi_\alpha \Phi_{total} / \varphi_\alpha - \Phi_{total, s_1 \cap s_2})^2 / \Phi_{total, s_1 \cup s_2}^4 , \qquad (17)$$

(see for notations Eqs. (5) and (6)). Therefore, the Jaccard coefficient defined by Eq. (14) can be conveniently normalized as

$$\iota_{s_1 \wedge s_2} = (J_{s_1 \wedge s_2} - <J_{s_1 \wedge s_2}>_{random}) / \sigma(J_{s_1 \wedge s_2}) . \qquad (18)$$

For random sequences, these normalized deviations are governed approximately by the Gaussian statistics. Below, the Jaccard coefficients (henceforth, JC) will always mean the corresponding normalized deviations.

The overlapping between TAMGI components $\Gamma_{s_1}$ and $\Gamma_{s_2}$ always provides the contribution into components $\Gamma_{s_1+s_2}$ and $\Gamma_{|s_1-s_2|}$. In other words, the overlapping components $\Gamma_{s_1}$ and $\Gamma_{s_2}$ may be considered as a partial source for generation of the components $\Gamma_{s_1+s_2}$ and $\Gamma_{|s_1-s_2|}$. The efficient generation needs both significant Jaccard correlations between components $\Gamma_{s_1}$ and $\Gamma_{s_2}$ as well as significant deviations defined by Eq. (7).

A set of JC is calculated at $s_1$ fixed and $s_2$ running from 1 to $S_2$, the value for $s_2 = s_1$ being discarded. The sets of JC provide an additional criterion for quasi-periodicity of motifs. In particular, if $s_1$ = $p$ and $s_2 = 2p$, the value of $\iota_{p \wedge 2p}$ should be high; or more symmetrically, $s_1 = 2p$, $s_2 = p$ and $s_2 = 3p$, while the values of $\iota_{2p \wedge p}$ and $\iota_{2p \wedge 3p}$ should both be high. For persistent quasi-periodicity, there is a series of high values for $\iota_{p \wedge kp}$, where $p$ is the period and $k = 2, 3, ...$ The multiplication and modification of repeated patterns is one of the molecular mechanisms responsible for the generation of RNAPS.

## 2.3. Statistical criteria and general remarks

The normalized deviations defined by Eqs. (7) and (18) unify the comparison of genomes with different lengths and nucleotide compositions. Primarily, we are interested in the most significant effects. For this reason, all data in Supplements are additionally presented in the ranked form. The related features for the values of top ranks are definitely non-random (the corresponding Gaussian probabilities are less than $10^{-10}$). The robustness of ranking with respect to mutations was assessed on limited sets of isolates. The corresponding variations in the normalized deviations were not higher than 0.5. Therefore, the ranked values were additionally grouped by grades (grade = 1) for the semi-quantitative assessment of divergence between these values. The values belong to the same grade if the absolute difference between them does not exceed 1; a value with difference between 1 and 2 with respect to a reference should be attributed to a lower grade etc.

As the main information in viral genomes is related to the coding for different proteins needed for virus proliferation, the signals corresponding to genome packaging are mainly evolved using the redundancy of the genetic code. The quasi-periodicity with period $p = 3$ appears to be the most pronounced in viral genomic sequences (for a review and further references see, e.g., Lobzin & Chechetkin, 2000; Marhon & Kremer, 2011). The features related to packaging and other molecular mechanisms are displayed as peaks in the sets of normalized deviations for the steps multiple to three.

## 3. Results

The encapsidation of coronaviral genomes is assumed to be performed immediately after the replication stage (Chang et al., 2014) in order to prevent viral RNA genomes from the action of host enzymes and from the damages during packaging of the genome within the membrane envelope. Similarly to the viruses with icosahedral capsids (Stockley et al., 2016), the encapsidation of viruses with helical capsids should also include more rapid assembly stage and subsequent slower packaging rearrangement of the RNA genome within the helical capsid. Taking into account the filament-like geometry of the helical capsid, the dynamics of the latter process should resemble the reptation of polymers (De Gennes, 1979). The corresponding assembly and packaging signals may generally be different (Chechetkin & Lobzin, 2019). Chang et al. (2009) found multiple (at least three) nucleic acid binding sites in N proteins that may be related to such a difference (see also Cubuk et al., 2021).

We restricted ourselves to the analysis of the genomes of alpha- and betacoronaviruses including all known pathogenic human coronaviruses. A batch of selected related genomes for animal coronaviruses was chosen for comparison. The grouping of the genomes is performed according to their evolutionary closeness. The first three groups belong to betacoronaviruses, whereas the fourth group comprises human, bat, pig, and camel alphacoronaviruses. For all human coronaviruses we used the standard reference genomic sequences from GenBank, whereas the accession numbers for the animal isolates will be presented separately in the text.

Strictly, each set of TAMGI motifs for a particular step $s$ should be considered as a correlational entity. The mutual overlapping of sets for different steps assessed by JC deviations provides information about their potentially coordinated action in different molecular mechanisms. As discussed above, the overlapping sets with steps $s_1$ and $s_2$ yield contribution into the sets with $s_1 + s_2$ and $|s_1 - s_2|$. The latter combinations will be called channels and provide insight into a partial mechanism of the motif generation in the network of overlapping motifs.

### 3.1. RNAPS in the genomes of SARS-related coronaviruses

We begin the study of RNAPS with the genomes of SARS-CoV (GenBank accession: NC_004718.3), SARS-CoV-2 (NC_045512.2) and SARS-related bat coronaviruses (accession numbers: DQ071615.1, DQ412043.1, DQ412042.1, MT782115.1, FJ588686.1, GQ153548.1, GQ153547.1, and DQ022305.2). The isolate with the accession FJ588686.1 (henceforth, SARS-Bat-CoV) was chosen for the presentation in the main text and Supplements. SARS-related coronaviruses belong to the lineage B of betacoronaviruses (Luk et al., 2019). Fig. 1 shows the normalized deviations defined by Eq. (7) for the steps $s$ within the interval 1–550 (left column) and the JC normalized deviations defined by Eq. (18) for the step $s_1 = 54$ and steps $s_2$ within the interval 1–600 (right column). As we are interested in the most pronounced effects, the dynamical range for the TAMGI and JC deviations here and below was restricted to approximately four grades (see Section 2.3).

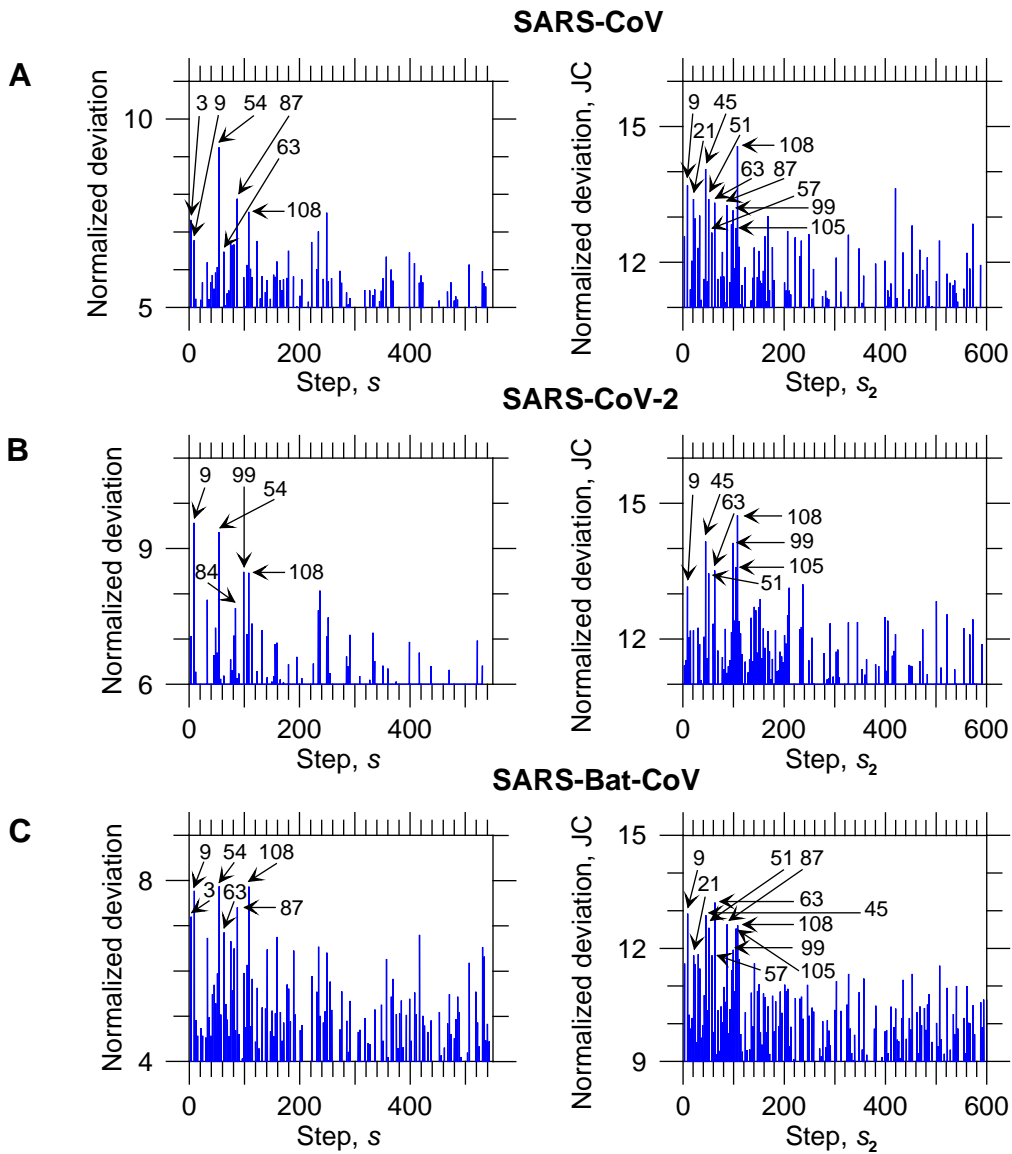

**Figure 1.** The spectra for the TAMGI (on the left) and JC (on the right) normalized deviations defined by Eqs. (7) and (18) for the genomes of the SARS-related coronaviruses. The TAMGI deviations were calculated within the range $s = 1–550$, whereas the deviations for JC were calculated for $s_1 = 54$ and $s_2 = 1–600$. **A**, SARS-CoV; **B**, SARS-CoV-2; **C**, SARS-Bat-CoV. The characteristic steps $s$ are explicitly marked by arrows.

The deviation $\kappa_s$ with $s = 54$ appeared to be the highest in the complete TAMGI spectrum for the genome of SARS-CoV and is more than a grade higher than the nearest ranked deviations ($\kappa_{54} = 9.25$, $\kappa_{87} = 7.89$, and $\kappa_{108} = 7.53$). The quasi-periodic character of the deviations $\kappa_{54 \times k}$ ($k = 1, 2, ...$) is clearly seen in Fig. 1A in accordance with the previous results (Chechetkin & Lobzin, 2020). The characteristic top ranked deviations are marked explicitly by arrows. The JC deviations shown on the right display the TAMGI motifs coordinated with those for $s = 54$. The motifs for the steps $s_2$ corresponding to the higher JC at $s_1 = 54$ can be called co-evolving to the motifs associated with $s_1 = 54$. As expected for quasi-periodic motifs with the period about $p = 54$, the highest Jaccard correlations are observed for $s_2 = 108$ (see Fig. 1A, right). The association with the octamer units corresponding to a half of period or $s_2 = 27$ is four grades lower, i.e. relatively weak. The characteristic values $s_2$ associated with the pronounced JC deviations are shown by arrows. As discussed in Section 2.2, the superimposed TAMGI motifs yield contribution into the generation of motifs corresponding to the sum and absolute difference of their steps. In particular, the motifs with $s_2 = 45$ and $s_2 = 9$ provide a contribution to the motifs with $s_1 = 54$ because 45+9=54. The same concerns the motifs with $s_2 = 63$ and $s_2 = 9$ because 63–9=54. The multiplication of the motifs with $s_2 = 9$ may also provide a contribution to the motifs with $s_1 = 54$, as 9×6=54. Other combinations such as 45+63, 21+87, 9+99 provide a contribution to the motifs with the doubled period, $2p = 108$. Modification of motifs during quasi-random molecular evolution makes the periodic patterns for $s = 54$ fuzzy. Indeed, significant correlations between patterns for $s_1 = 54$ and $s_2 = 51$ or 57 as well as the correlations with $s_2 = 105$ are also seen in Fig. 1A, right. The comparison with the cryo-EM data (Chang et al., 2014) indicates that the motifs for $s = 54$ should be associated with the packaging signals, whereas the motifs for $s = 51, 57, 105, 45, 9$ (via 45+9 or 9×6), 87, 21, and 99 can be associated in part with the assembly signals. Such multi-channel participation of the assembly signals ensures the higher rate of encapsidation.

In the genome of SARS-CoV-2 the deviations $\kappa_s$ with the steps $s = 9$ and 54 are the highest and belong to the same grade ($\kappa_9 = 9.57$ and $\kappa_{54} = 9.37$). The next nearest deviations correspond to the steps $s = 99$ and 108 ($\kappa_{99} = 8.48$ and $\kappa_{108} = 8.47$). Unlike the SARS-CoV genome with the high deviation for $s = 87$ and the deviation for $s = 84$ lower by three grades, the deviation for $s = 84$ was among top ranked for the genome of SARS-CoV-2 and a grade higher in comparison with the deviation for $s = 87$ (Fig. 1B, left). The JC deviation for $s_2 = 108$ (Fig. 1B, right) is again the highest that proves the quasi-periodic character of the motifs with $s = 54$. The motifs associated with the combinations 45+9, 99+9, 99–45, 63–9, 45+63 and the fuzzy motifs with $s = 51$ and 105 may be associated with the assembly signals as discussed above.

For a batch of SARS-related bat genomes, the deviations for the steps $s = 54, 108, 87, 9$, and 3 were persistently among top ranked and belonged commonly to the same grade (except DQ412042.1, for which the deviation for $s = 9$ was a grade higher in comparison with the others). The ranking for these steps may permute depending on a particular isolate. These features are also typical of the chosen isolate FJ588686.1 ($\kappa_s = 7.87, 7.87, 7.77, 7.41$, and 7.20 for $s = 54, 108, 9, 87$, and 3, respectively; see also Fig. 1C, left). Though the JC deviation for $s_2 = 108$ was not the highest, the JC deviations for $s_2 = 63, 9, 45, 87, 108, 51$, and 105 belonged to the same top ranked grade (Fig. 1C, right), providing the combinations associated with the nucleocapsid assembly similarly to the genome of SARS-CoV.

Figs. 1A–C reveal several reproducible features. The deviations for $s = 54$ and 108 were invariably among the highest, while the higher TAMGI and JC deviations were approximately clustered within the range $s \leq 200$. The corresponding sets for the JC deviations were also calculated for the steps $s_1 = 21, 24, 27, 30, 42, 48, 51, 57, 60$, and 63 and the steps $s_2$ running from 1 to 600. These data are collected in Supplement S1 together with TAMGI deviations. The JC deviations at these steps $s_1$ were invariably among the top ranked for $s_2 = 54$, indicating the involvement of packaging periodicity $p = 54$ into various regulatory mechanisms and multi-channel generation of RNAPS.

The distribution of motifs over the genome was calculated using windows of width $w = 216$ and sliding step of 108 (half-overlapping windows). For the human and bat SARS-CoV the TAMGI motifs corresponded to $s = 9, 54$, and 87, whereas for SARS-CoV-2 the step $s = 87$ was replaced by the more typical $s = 84$. The profiles for the distribution of motifs over the genome were calculated both for all $k$-mers and separately for $k$-mers with $k \geq 5$. The resulting profiles are shown in Fig. 2. The Pearson correlations between counterpart profiles for all $k$-mers and for $k$-mers with $k \geq 5$ were about 0.6–0.7. The corresponding peaks and troughs on the profiles may indicate the involvement of related motifs in the particular regulation mechanisms (cf. the comparison of such features with the experimentally verified examples below). Interestingly, the profile for $k \geq 5$ and $s = 9$ revealed the highest peak in the region of the gene coding for N protein in the genome of SARS-CoV-2 (windows #267 and 268 in Fig. 2B). The binding of N proteins with this region might autocontrol the expression of the gene coding for N protein via a feedback loop. The profiles for all three lineage B coronaviruses display pronounced peaks within region of the gene coding for S protein (windows #200–220 in Fig. 2). The detailed data related to the profiles are collected in Supplement S2.

All $k$-mer motifs with $k \geq 6$ corresponding to $s = 9, 54$, and 87 for the human and bat SARS-CoV and to $s = 9, 54$, and 84 for SARS-CoV-2 are explicitly reproduced in Supplement S3. We found that the motifs for $s = 54$, $k = 9$ contained transcription regulatory sequences ACGAAC typical of SARS-related coronaviruses (Woo et al., 2010) in all three genomes. Such regulatory sequences were contained within motifs TACGAACTT at the similar start positions on the genomes (site on the genome counted by the position of the first nucleotide in a motif), 26108 (close to the start at 26117 of the gene coding for E protein in the genome of SARS-CoV), 26236 (close to the start at 26245 of the gene coding for E protein in the genome of SARS-CoV-2), and 25469 (close to the start at 25478 of the gene coding for E protein in the genome of SARS-Bat-CoV). Besides, the motif CTAAACAT with *k* = 8 and *s* = 9 positioned at 16049 on the genome of SARS-CoV-2 (upstream from the gene coding for nsp13) contained transcription regulatory sequence CTAAAC typical of alphacoronaviruses and of lineage A betacoronaviruses (Woo et al., 2010).

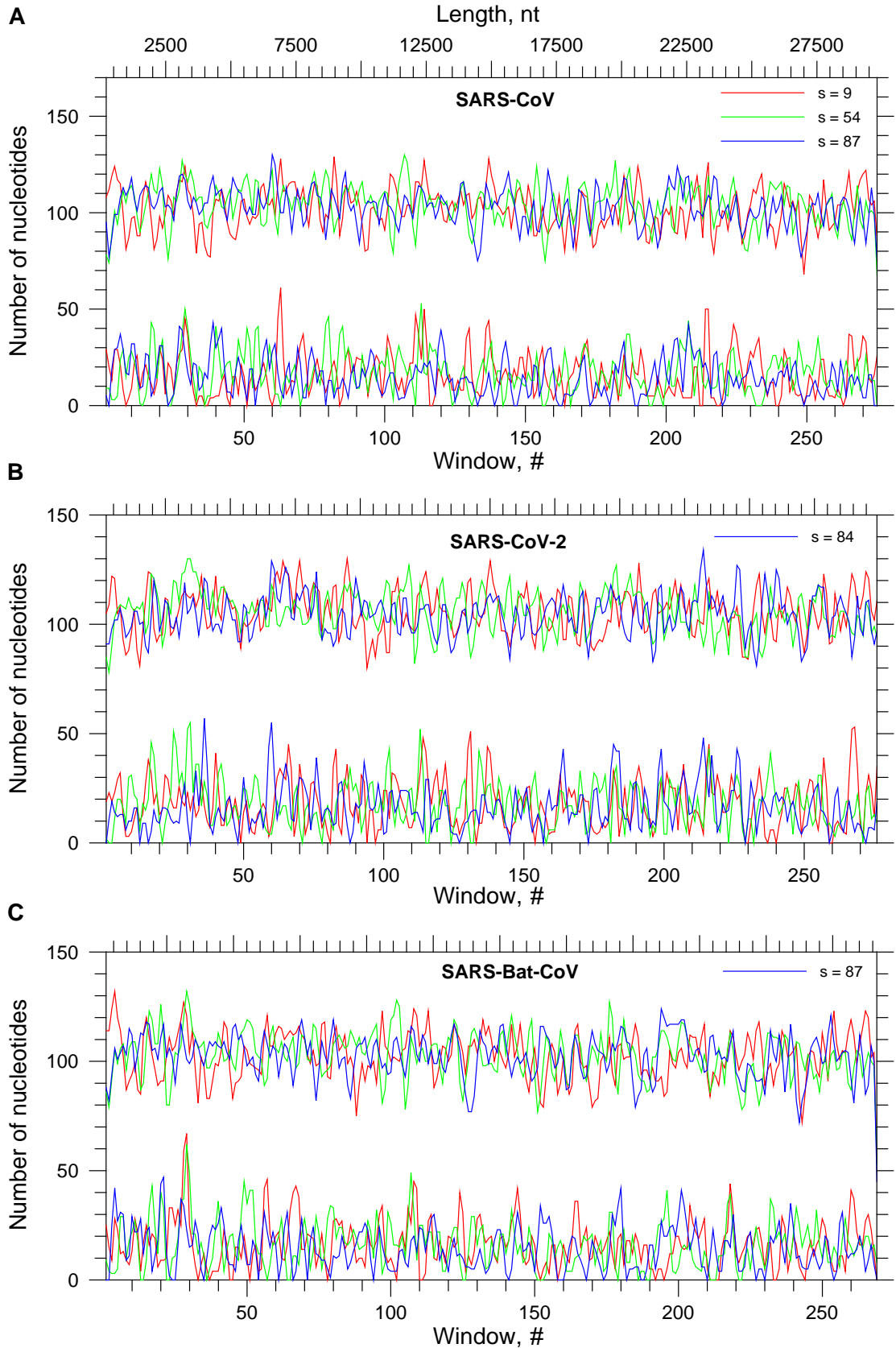

**Figure 2.** The distributions of motifs over the genomes of SARS-CoV, SARS-CoV-2 and SARS-Bat-CoV calculated with windows of width $w = 216$ and sliding step of 108 (half-overlapping windows). The corresponding TAMGI steps are shown in the inserts. The upper profiles correspond to all motifs ($k \geq 1$), whereas the lower profiles correspond to $k \geq 5$.

The relationship between particular RNAPS and transcription regulatory sequences corresponds to the experimentally established multifunctional role of N proteins which participate not only in the assembly/packaging of the ribonucleocapsid but also in the regulation of the replication-transcription processes (Grossoehme et al., 2009; Hurst et al., 2010, 2013; McBride et al., 2014; Verheije et al., 2010; Yang et al., 2021).

The complete RNAPS repeats with $k \geq 6$ were rather rare in the genomes of particular SARS-related coronaviruses (about 10–15 pairs of repeats from the total number of motifs about 200 at the each step $s$). The numbers of completely coincident RNAPS with $k \geq 6$ for the genomes of human and bat SARS-CoV with the coordinated positioning of motifs on both genomes amounted to 112 from 205 motifs at $s = 9$ and to 102 from 216 motifs at $s = 54$ (the total number of motifs refers to the genome of SARS-CoV). The lengths of coincident motifs $k$ varied from 6 up to 13. Such close correspondence indicates clear evolutionary conservation of the evolved RNAPS. The number of completely coincident RNAPS with $k \geq 6$ in the genomes of SARS-CoV and SARS-CoV-2 was much lower and amounts to 42 at $s = 9$ and to 38 at $s = 54$ with much variable positioning. Nevertheless, the positioning of the longer coincident RNAPS with $k \geq 7$ was close as well, e.g., TTTGACC (with the start sites, respectively, at 12166 and 12236); TTGATAA (14797, 14867); TTATGGG (15211, 15281); ACAGATG (27070, 27198); GTGGTAGA (9197, 9267); ACTTCTTT (26167, 26295); TGATTCTGA (12108, 12178); ACAATGTTG (14578, 14648); TACGAACTT (26108, 26236); GAATGTGGCTAA (12147, 12217) at $s = 54$. These listed RNAPS coincided in all three genomes including their close positioning on the genomes. Similar triple coincidences were obtained for the motifs at $s = 9$, ATGAATC (14994, 15064, 14396); TTACCCA (15860, 15930, 15262); CAACAATG (12342, 12412, 11744); ATGACTTT (14629, 14699, 14031); TGATTGTT (14810, 14880, 14212); CATGCTTA (15254, 15324, 14656); TGGTATTC (26187, 26315, 25548); and AACAACAAG (28841, 28992, 28241). We also studied motifs coincident up to one mismatch for $k = 6$ and $s = 9, 54$ (Supplement S3). The sums of complete and one-mismatch coincidences for the pairs SARS-CoV–SARS-CoV-2 and SARS-CoV–SARS-Bat-CoV were close despite the differences within complete and one-mismatch groups.

### 3.2. RNAPS in the genomes of MERS-related coronaviruses

MERS-CoV discovered in 2012 is responsible for acute respiratory syndrome in humans with overall mortality around 35.7% (Azhar et al., 2019). Though bats and alpacas are considered to be the potential reservoirs for MERS-CoV, dromedary camels seem to be the only animal host responsible for the camel-to-human virus transmission and spread of human infections (Azhar et al., 2014; Chan et al., 2015; Mohd et al., 2016; Omrani et al., 2015). This endemic disease was found in dromedary camel populations of East Africa and the Middle East. Human and camel MERS-CoV belong to the lineage C of betacoronaviruses (Luk et al., 2019). Despite the divergence of amino acid sequences, N protein for MERS-CoV retains the main structural similarity with N proteins for the other coronaviruses (Chang et al., 2014; Nguyen et al., 2019; Peng et al., 2020).

In this section we study RNAPS in the genomes of MERS-related coronaviruses. We used the reference sequence for human MERS-CoV (accession: NC_019843.3) and a limited batch of camel MERS-CoV genomes (KF917527.1, KT368860.1, and KT368890.1). The isolate KT368860.1 (henceforth, MERS-Camel-CoV) highly homologous (about 99%) to human MERS-CoV was chosen for presentation. The TAMGI deviations for the human and camel MERS-CoV in the range of the steps $s$ from 1 to 550 are shown in Fig. 3. The top ranked grade comprises nearly 30 various steps (see Supplement S1). The highest deviations correspond to $\kappa_{171}$ = 7.09 and 7.03 (for human and camel MERS-CoV, respectively); $\kappa_{42}$ = 7.03 and 6.93; as well as $\kappa_{63}$ = 6.77 and 6.91, whereas the deviations putatively associated with RNAPS were $\kappa_{108}$ = 6.11 and 6.35 (ranking positions 25 and 14, respectively); and $\kappa_{54}$ = 5.79 and 5.92 (38 and 34). The JC spectra for $s_1$ = 54, 21, 42, and 63 and $s_2$ from 1 to 600 are shown in Fig. 4 only for human MERS-CoV; the counterpart spectra for MERS-Camel-CoV were close (see also Supplement S1). The combined analysis of TAMGI and JC deviations revealed that the RNAPS were coordinated with the main channels corresponding to the doubled or approximately doubled periods, 171–63 → 108 and 42 + 63 → 105. The respective Jaccard correlations between motifs for $s_1$ = 54 and for $s_2$ = 105 were higher than those between motifs for $s_1$ = 54 and for $s_2$ = 108, $\iota_{54\wedge105}$ = 12.10 (ranked position 4); and $\iota_{54\wedge108}$ = 11.50 (position 23). The JC value $\iota_{54\wedge171}$ = 12.12 corresponded to the ranked position 2, whereas the top ranked value, $\iota_{54\wedge96}$ = 13.59, was a grade higher. The channels 96–42, 99+9, 96+9, 93+15, 36+69 may be associated with the assembly signals. Nearly all these channels correspond to the doubled period.

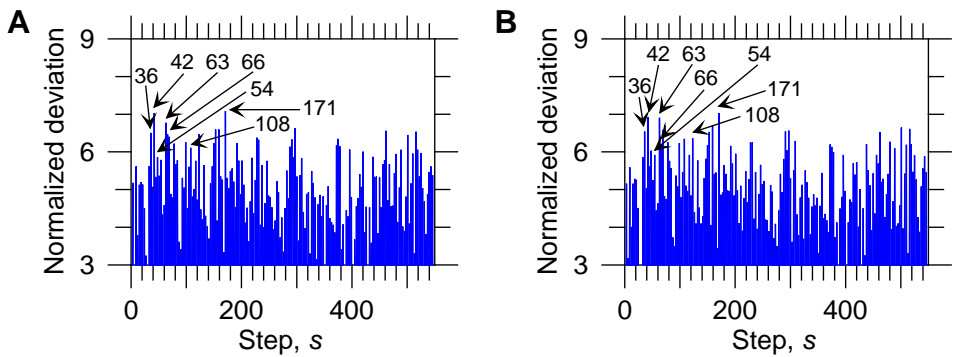

**Figure 3.** The spectra for the TAMGI normalized deviations defined by Eq. (7) for the genomes of the MERS-related coronaviruses. The TAMGI deviations were calculated within the range $s$ = 1–550. **A**, MERS-CoV; **B**, MERS-Camel-CoV. The characteristic steps $s$ are explicitly marked by arrows.

The motifs associated with $s$ = 21 can be attributed to quasi-periodic ones. The correlations between motifs for $s$ = 42 with the motifs for $s$ = 21 and $s$ = 63 are, respectively, one and two grades lower than that for $s$ = 84 (Fig. 4C), indicating approximately independent quasi-periodicity for $s$ = 42. The correlations $\iota_{21\wedge171}$ = 13.19 belonged to the top ranked for $s_1$ = 21 (see Fig. 4B and Supplement S1).

**Figure 4.** The spectra for the JC normalized deviations defined by Eq. (18) for the genome of MERS-CoV. The deviations for JC were calculated for $s_2 = 1$–600 and **(A)** $s_1 = 54$; **(B)** $s_1 = 21$; **(C)** $s_1 = 42$; and **(D)** $s_1 = 63$. The characteristic steps $s_2$ are explicitly marked by arrows.

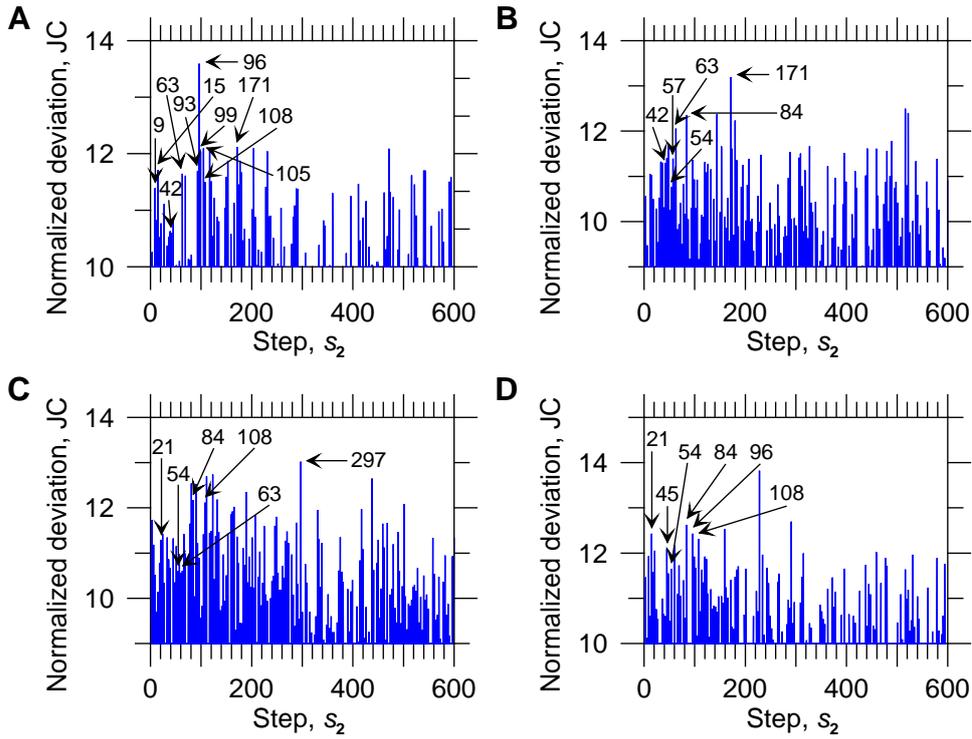

The profiles for the motifs for $s$ = 42, 54, 63, 105, 108, and 171 were calculated using windows $w$ = 216 with the sliding step 108 as described above. The corresponding profiles for human MERS-CoV are shown in Fig. 5. Hsin et al. (2018) established experimentally that a fragment on the genome of human MERS-CoV occupying sites 19712–19969 can be identified to function as a packaging signal dependent on N proteins. Such packaging signal is responsible for the transport of the RNA genome into the membrane envelope. This fragment corresponds to the windows #183 and #184. We detected within these windows the highest peak over complete profile for the motifs $s$ = 42, whereas all other motifs were depleted in this region (see Fig. 5). For the next downstream window #185, we found very high local peak for the motifs with $s$ = 171 especially pronounced for the longer motifs with $k \geq 5$ (see Fig. 5B). Note also the high peaks for the profile $s$ = 105 in the vicinity of the start region of the gene coding for N protein (window #265) and for the profiles $s$ = 42 and 108 in the vicinity of the end region for this gene (window #274).

All motifs with $k \geq 6$ for $s$ = 42, 54, 63, 105, 108, and 171 are collected in Supplement S3. We found the following motifs within region of packaging signal 19712–19969: $s$ = 42, $k$ = 6, ACCACT (start at 19712); CTGTAC (19758); ACATTT (19806); TCATGT (19956); $k$ = 7, CGATTTC (19774); TATATGT (19915); $k$ = 8, GTGTATGT (19872); $k$ = 9, ATATTTATG (19848); $s$ = 54, $k$ = 8, ATGCTAAG (19746); $s$ = 63, $k$ = 7, TGTAAGT (19877); $s$ = 105, $k$ = 6, TGAACT (19738); AATGGT (19937); $k$ = 8, TACTGCTA (19858); $s$ = 108, $k$ = 6, TGCTGT (19756); $k$ = 7, TATGCTA (19745); $s$ =

171, $k = 7$, AATGCCA (19970); $k = 8$, CACTGATA (19885); TGATGTTA (19894). The positioning of these motifs within packaging signal specifies RNA-protein interactions, the motifs ACCACT (19712) and AATGCCA (19970) being located strictly at the start and the end of the packaging signal. Note also motif $s = 42$, $k = 9$, CCCTTTGTC (27596) nearby start of the gene coding for E protein (27590) which may affect the transcription of this gene.

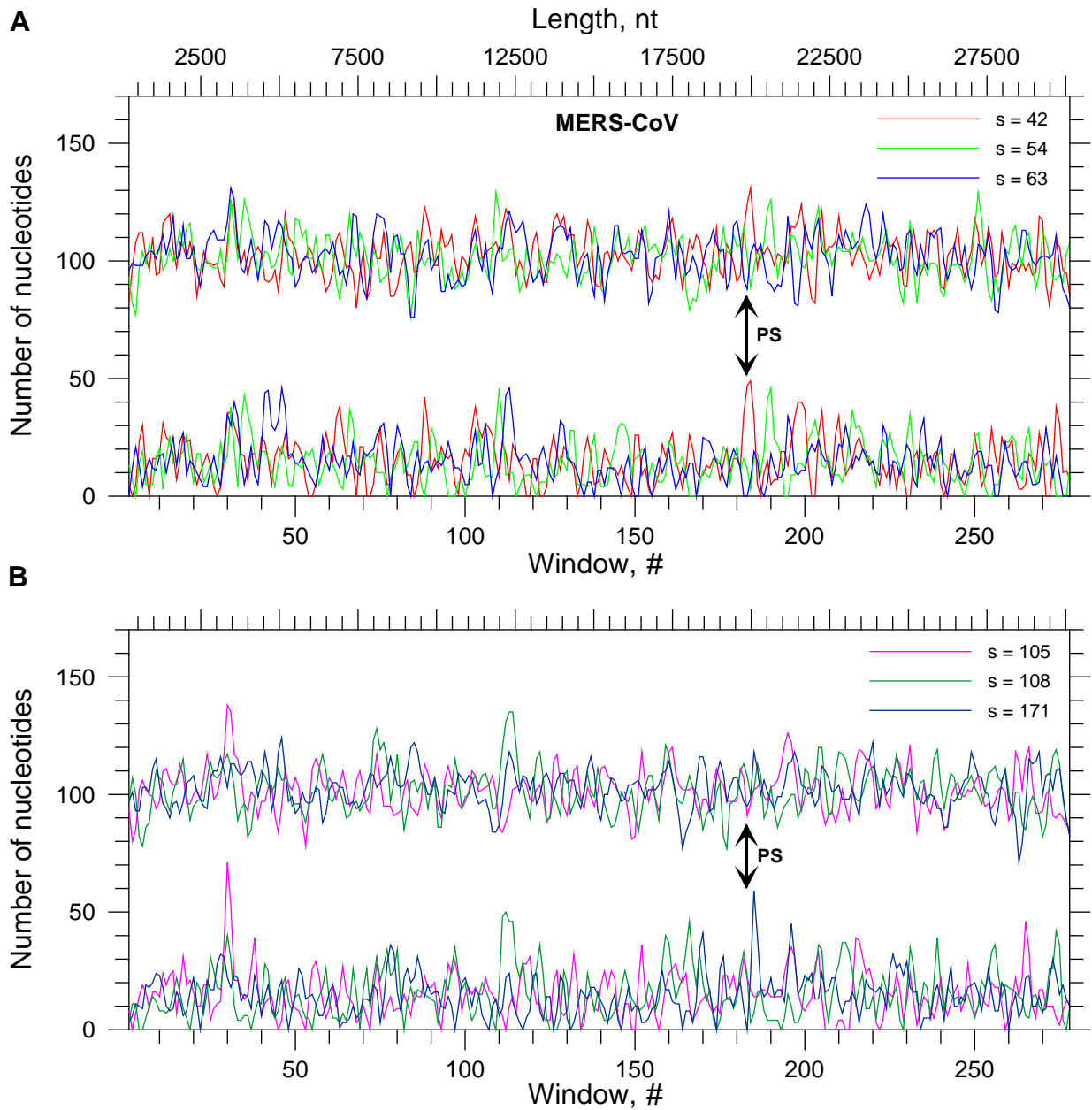

**Figure 5.** The distributions of motifs over the genome of MERS-CoV calculated with windows of width $w = 216$ and sliding step of 108 (half-overlapping windows). The corresponding TAMGI steps are shown in the inserts. The upper profiles correspond to all motifs ($k \geq 1$), whereas the lower profiles correspond to $k \geq 5$. The two-sided arrows indicate the middle of the experimentally verified packaging signal located at the sites 19712–19969.

### *3.3. RNAPS in the genomes of HCoV-OC43, HCoV-HKU1 and MHV*

In this section we study RNAPS in the genomes of HCoV-OC43, HCoV-HKU1 and MHV. All these viruses belong to the lineage A of betacoronaviruses (Woo et al., 2010; Luk et al., 2019). If the lengths of the genome and of the gene coding for N protein for SARS-CoV are taken as a reference ($M = 29751$, $L_N = 1269$), their counterparts are distinctly longer for HCoV-OC43, HCoV-HKU1 and MHV ($M = 30741$, $L_N = 1347$; $M = 29926$, $L_N = 1326$; and $M = 31357$, $L_N = 1365$; respectively). We chose for MHV the standard reference sequence for the strain MHV-A59 C12 mutant (accession: NC_001846.1); for HCoV-OC43 and HCoV-HKU1 we also used the reference sequences (NC_006213.1 and NC_006577.2). The therapeutic targeting of N protein for the pathogenic coronavirus HCoV-OC43 showed how the knowledge of molecular mechanisms of encapsidation can be used in medical applications. It was found that a compound known as PJ34 inhibits coronavirus replication and the RNA-binding activity of HCoV-OC43 N protein (Chang et al., 2016).

Gui et al. (2017) found that parameters for the helical nucleocapsid of MHV were close to that of SARS-CoV (up to a slight uncertainty in the helix pitch). The structure of N protein for HCoV-OC43 was also close to that for SARS-CoV and MHV up to several divergent features (Chang et al., 2014; Peng et al., 2020). Taking into account the lengths of the genomes and of the genes coding for N proteins, a slight shift from $p = 54$ to $p = 57$ may be expected in the main periods for RNAPS in the genomes of HCoV-OC43, HCoV-HKU1 and MHV. Testing this hypothesis was among the matters of the study in this section.

The related spectra for TAMGI deviations for $s = 1$–550 and JC deviations for $s_1 = 54$ and $s_2 = 1$–600 are shown in Fig. 6 (on the left and right, respectively). The spectra for JC deviations with $s_1$ corresponding to the highest TAMGI deviations, $s_1 = 21$ (HCoV-OC43) and $s_1 = 24$ (HCoV-HKU1 and MHV) as well as the JC spectra for $s_1 = 57$ (all three viruses) are shown in Fig. 7 (on the left and right, respectively). The main characteristics for the genomes of these viruses were as follows.

The highest deviation $\kappa_{21} = 10.16$ for the genome of HCoV-OC43 was a grade above the two nearest top ranked deviations, $\kappa_{66} = 9.00$ and $\kappa_3 = 8.67$. The deviation $\kappa_{108} = 8.36$ was at the ranked position 6 (in comparison with $\kappa_{114} = 6.60$ at the position 78), whereas $\kappa_{57} = 7.77$ and $\kappa_{54} = 7.48$ were at the positions 22 and 31, respectively (Fig. 6A, left and Supplement S1). The correlations $\iota_{54\wedge 9} = 15.95$ were the highest for $s_1 = 54$, whereas $\iota_{54\wedge 105} = 14.76$ corresponded to the ranked position 5. The correlations $\iota_{57\wedge 102} = 15.51$ and $\iota_{57\wedge 108} = 15.08$ were on the positions 2 and 3, respectively, and were more than a grade higher than the correlations with $s_2 = 114$ ($\iota_{57\wedge 114} = 13.84$). The combined analysis of TAMGI and JC spectra indicates that the fuzzy periodicity $p = 54$ is coordinated mainly with the channels 9×6, 33+21, 63–9 and the doubled periods 105, 9+96, 33+75. The motifs for $s = 21$ may also be attributed to quasi-periodic ones and yield contributions into the generation of motifs with $s = 54$ and 57 (Fig. 7A).

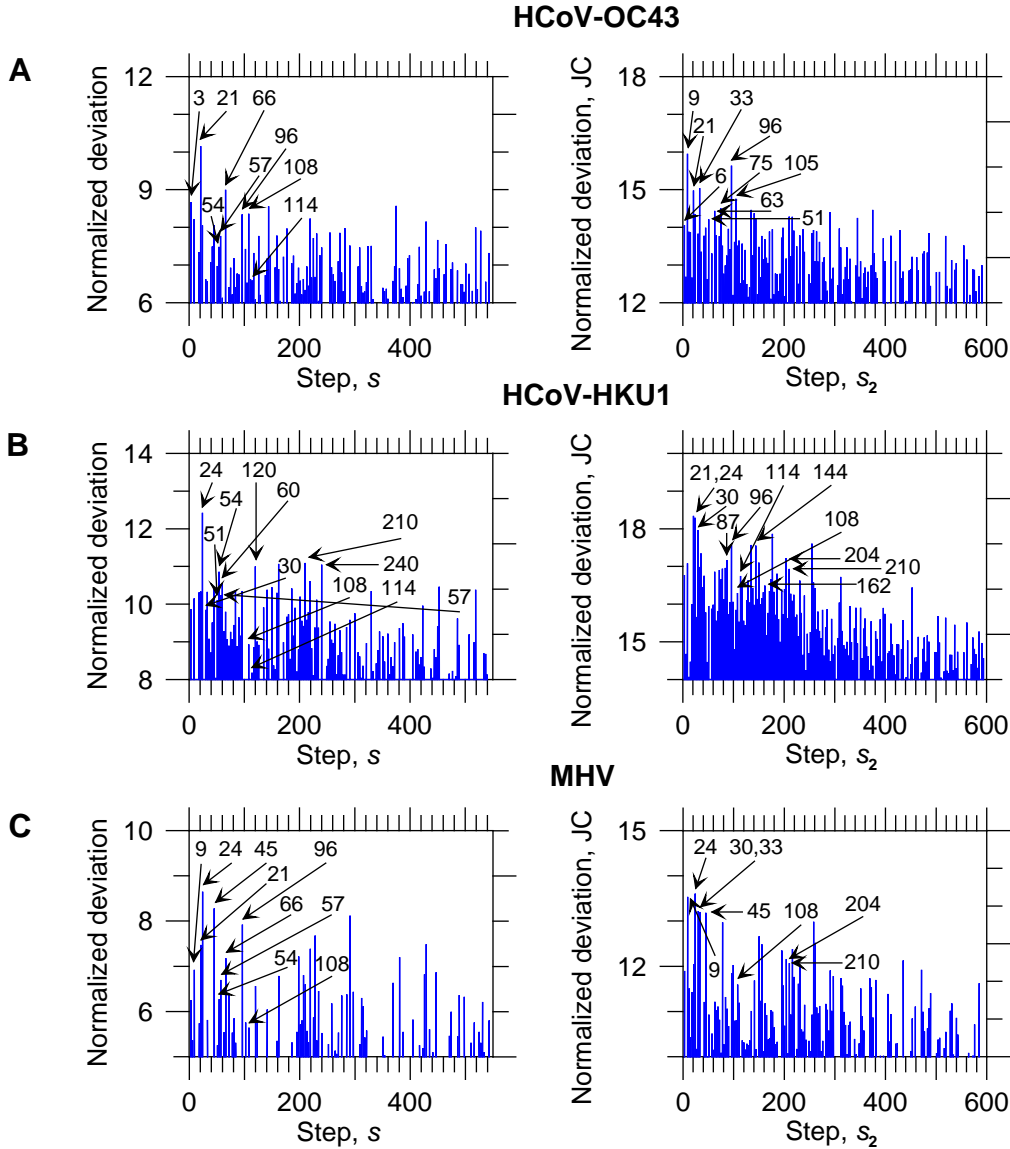

**Figure 6.** The spectra for the TAMGI (on the left) and JC (on the right) normalized deviations defined by Eqs. (7) and (18) for the genomes of the lineage A betacoronaviruses. The TAMGI deviations were calculated within the range $s = 1$–550, whereas the deviations for JC were calculated for $s_1 = 54$ and $s_2 = 1$–600. **A**, HCoV-OC43; **B**, HCoV-HKU1; **C**, MHV. The characteristic steps $s$ are explicitly marked by arrows.

The genome of HCoV-HKU1 contains a fragment with the complete tandem repeats of 30 nt coding for amino acids NDDEDVVTGD (Woo et al., 2006). Such a feature is very rare in viral genomes. In the reference genome, the complete tandem repeats occupy the sites from 3038 to 3460 and in the partly incomplete form persist up to 3517 within the gene coding for nsp3 (sites 2633–8719 within ORF1ab). Strongly modified by mutations, such repeats are scattered over the genome of HCoV-HKU1 and form larger correlational units as is seen from TAMGI deviations, $\kappa_{210} = 11.10$ (ranked position 2); $\kappa_{240} = 11.05$ (4); $\kappa_{120} = 11.00$ (5); $\kappa_{60} = 10.56$ (8) etc. Remarkably, the deviation for $s = 30$, $\kappa_{30} = 9.85$ appears to be only at the position 31. The highest deviation $\kappa_{24} = 12.42$ was a grade higher than the nearest deviation for $s = 210$. The deviations around RNAPS periodicity $p = 54$ were $\kappa_{51} = 10.11$ (26); $\kappa_{54} = 10.86$ (6); and $\kappa_{57} = 10.24$ (22), whereas the respective deviations for the doubled periods were two grades lower (with the highest deviation for $s = 108$). The TAMGI deviations are presented in Fig. 6B (left) and Supplement S1. The JC spectrum for $s_1 = 30$ reveals the highest correlations with $s_2 = 54$, $\iota_{30 \wedge 54}$

= 17.97 (in comparison with $\iota_{30^{\wedge}120}$ = 17.79 and $\iota_{30^{\wedge}60}$ = 17.67 at the ranked positions 2 and 3, respectively). The top three ranked JC for $s_1 = 54$ were with $s_2 = 21, 24$, and 30 (with respective JC deviations 18.34, 18.29 and 17.97) (see also Fig. 6B, right). The main channels related to the quasi-periodic RNAPS motifs can be associated with 24+30 and 21+30.

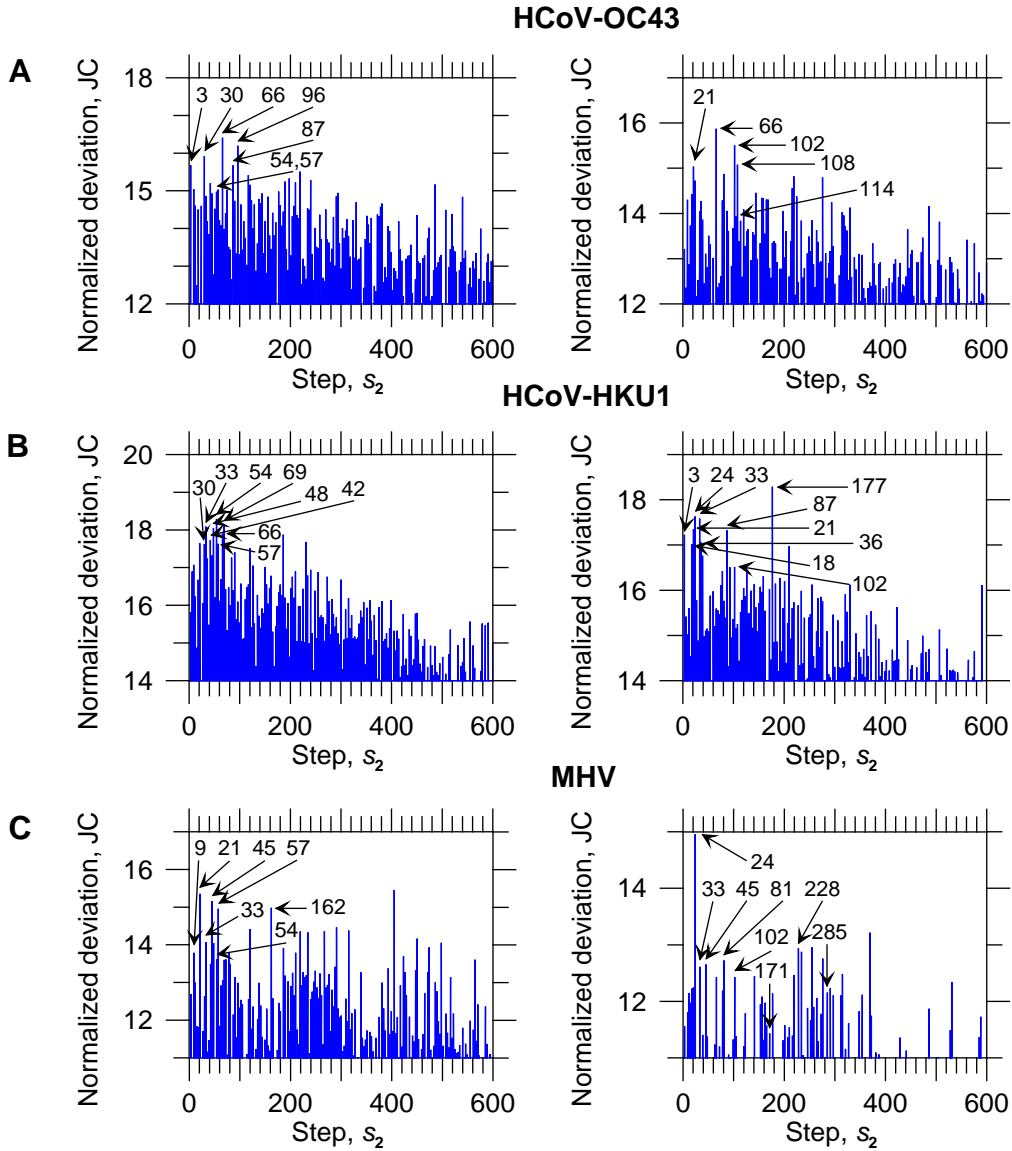

**Figure 7.** The spectra for the JC normalized deviations defined by Eq. (18) for the genomes of HCoV-OC43 **(A)**; HCoV-HKU1 **(B)**; and MHV **(C)**. The deviations for JC were calculated for $s_2 = 1–600$ and **(A)** $s_1 = 21$ (left) and 57 (right); **(B, C)** $s_1 = 24$ (left) and 57 (right). The characteristic steps $s_2$ are explicitly marked by arrows.

MHV is one of the best-studied coronaviruses. The top five ranked TAMGI deviations were $\kappa_{24}$ = 8.65; $\kappa_{45}$ = 8.28; $\kappa_{291}$ = 8.12; $\kappa_{96}$ = 7.92; and $\kappa_{228}$ =7.68 (see also Fig. 6C, left and Supplement S1). The steps $s = 96$ and 228 are the multiples of 4×24 and 4×57. The deviations for $s = 24$ can be attributed to quasi-periodic motifs. The deviations associated with RNAPS were $\kappa_{57}$ = 6.70 (position 17) and $\kappa_{54}$ = 6.27 (29), while $\kappa_{51}$ was a grade lower. The deviations for the doubled steps were $\kappa_{108}$ = 5.64 and $\kappa_{114}$ = 3.88. The top five JC deviations $\iota_{54^{\wedge}s_2}$ corresponded to $s_2 = 24, 9, 30, 33, 45$ and belonged to the same grade (Fig. 6C, right). The correlations $\iota_{57^{\wedge}s_2}$ were top ranked for $s_2 = 24$ and were a grade higher in comparison with $\iota_{54 \wedge 24}$. The channels related to $s = 54$ may be attributed to 9×6, 24+30, 9+45, whereas those for $s = 57$ may be attributed to 24+33, 81–24. In both cases the correlations $\iota_{s_1 \wedge s_2}$ for $s_1 = 54$ and 57 were higher for $s_2 = s_1 \times k$ with $k \geq 3$ rather than for $k = 2$ (with an approximate exception $s_2 = 102$ for $s_1 = 57$) (Fig. 7C and Supplement S1). The TAMGI spectrum for MHV strain ML-10 (AF208067.1) was nearly coincident with MHV-A59 for the top rank deviations (data not shown).

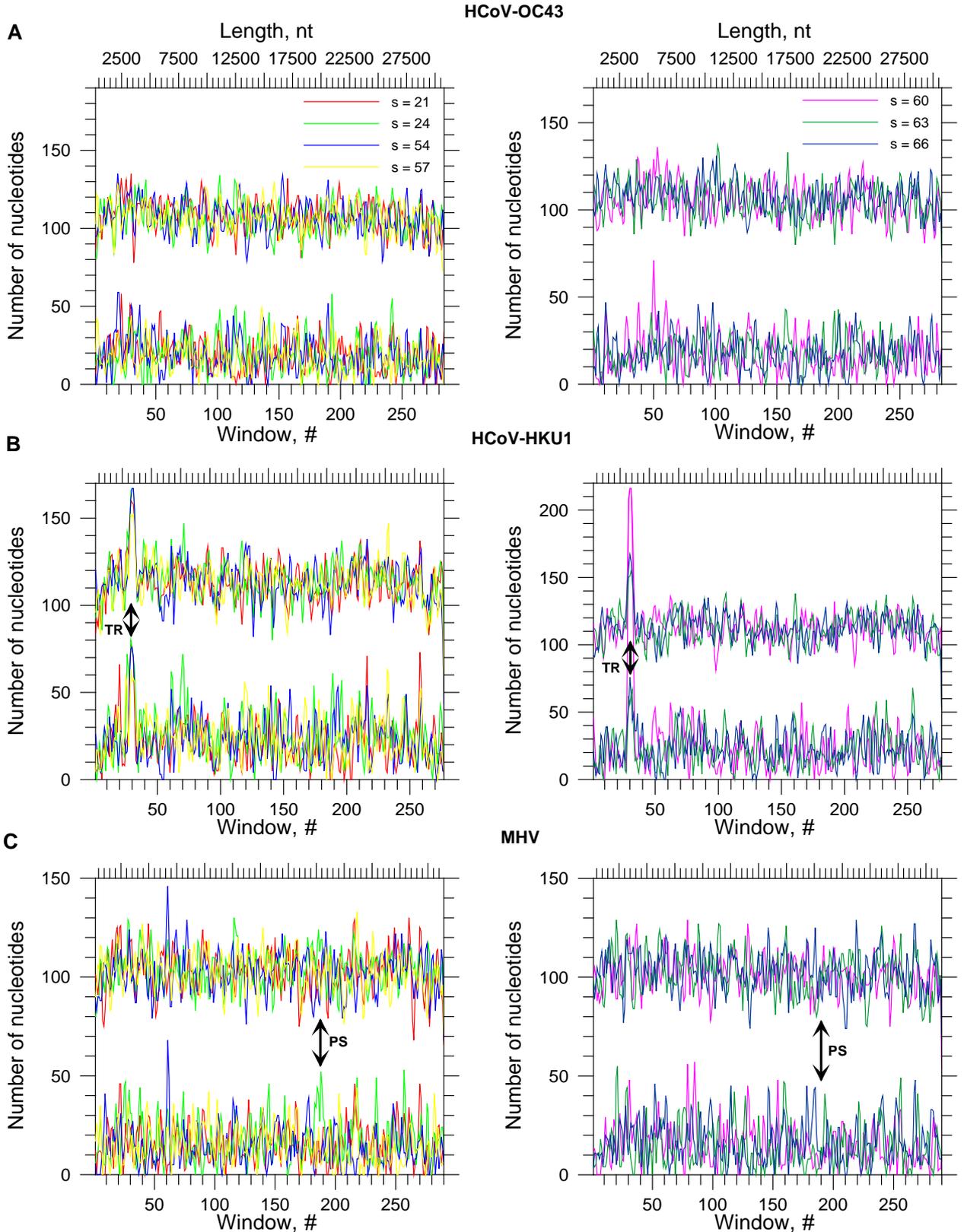

**Figure 8.** The distributions of motifs over the genomes of HCoV-OC43 **(A)**, HCoV-HKU1 **(B)** and MHV **(C)** calculated with windows of width $w = 216$ and sliding step of 108 (half-overlapping windows). The corresponding TAMGI steps are shown in the inserts. The upper profiles correspond to all motifs ($k \geq 1$), whereas the lower profiles correspond to $k \geq 5$. The two-sided arrows for HCoV-HKU1 **(B)** indicate the middle of a region with tandem repeats, whereas the arrows for MHV **(C)** indicate the middle of the experimentally verified packaging signal located at the sites 20273–20416.

Profiles for the motif distributions over the genome were calculated for half-overlapping windows of width $w = 216$ and the TAMGI steps $s = 21, 24, 54, 57, 60, 63, 66$. These data are presented in Fig. 8 and Supplement S2. We begin with the profiles for MHV because in this case there are experimentally verified data on the packaging signal (Fosmire et al., 1992; Kuo & Masters 2016; Masters, 2019; Narayanan, & Makino, 2001). Fosmire et al. (1992) established that a region of about 190 nt within the gene coding for nsp15 functions as a packaging signal with the minimal core fragment of 61 nt occupying the sites 20356–20416. Based on the bulged stem-loop RNA secondary structure, Kuo & Masters (2016) suggested the region 20273–20367 for the packaging signal. We will use the region 20273–20416 as a reference because the experiments indicate that broader regions function more efficiently. This region corresponds to the windows #188 and #189. As is seen from Fig. 8C and Supplement S2, there is a peak for the motifs $s = 24$ in the region with the packaging signal, whereas all other motifs were depleted in this region. The content of the longer motifs with $k \geq 5$ for $s = 24$ attained the absolute maximum in the region of the packaging signal. A similar yet lower peak for $s = 24$ and less stringent depletion of the other motifs were observed for the windows #185 and #186 upstream from the packaging signal. We also found the abundance of motifs with $s = 66$ for the windows #184 and #185. The abundance of one motifs and depletion of the others was also inherent to the packaging signal for MERS-CoV (Section 3.2).

The motifs with $k \geq 6$ for $s = 21, 24, 54, 57, 60, 63$, and 66 for the genomes of all three coronaviruses are collected in Supplement S3. We found the following motifs with $k \geq 6$ in the region 20273–20416 for the MHV genome: none for $s = 21, 54, 57, 60, 63$; $s = 24$, GGGAGCCTT (20285); GGAGCC (20310); AAGGTAATC (20319); GGTAATC (20333); GGTGATC (20345); GCGGTAAT (20355); and $s = 66$, GGTGGTA (20330). Note the reproducibility of GGAGCC in the first two motifs and GGTAAT in the next four motifs for $s = 24$ and the approximately phased positioning of these motifs.

A part of high peaks on the motif profiles is located in the region of the genes coding for the structural and accessory proteins. In particular, the high peaks for MHV, $s = 24, 63$ were located at the end region of the gene coding for S protein (windows #256 and 257), for $s = 66$ (windows #267 and 268, gene E), and $s = 21$ (window #270, gene M). The high peaks for the HCoV-HKU1 profiles were detected for $s = 21, 54$, and 57 within the region of the gene coding for M protein (windows #258 and 259), $s = 63$ within the region of the gene coding for E protein (windows #254 and 255), and $s = 21, 54$ within the region of the gene coding for S protein (window #216). The high peak for the HCoV-OC43 profiles was detected for $s = 24$ within the region of the gene coding for S protein (windows #241 and 242). Thus, the peaks on the profiles within the region of the gene coding for S protein can be considered as a common feature for all these lineage A coronaviruses. A possible relationship of this feature with the recombination is discussed in Section 4.3 below.

The profiles for motifs in the genome of HCoV-HKU1 reveal high peaks for all steps $s$ within the windows #29–32 corresponding to the tandem repeats (Fig. 8B). Generally, tandem repeats always are simultaneously a source of correlational TAMGI motifs which may participate in various molecular mechanisms. In particular, this region may contain RNAPS as well. The local peaks in this region were also found for MHV, $s = 24, 60, 63$ (Fig. 8C) and for HCoV-OC43, $s = 21, 24, 54, 57$ (Fig. 8A).

The local peak for the profile $s = 24$ in the case of HCoV-HKU1 was found within the same windows #188 and #189 as for the MHV packaging signal (Fig. 8B). The counterpart motifs for $s = 24$ were for HCoV-HKU1 at the phased positions close to those of MHV and contained the motif GGTAAT, GGTAATCT (20329); TGGTAAT (20340); GGTAAT (20353); TTGGTAAT (20363). The motifs for the other steps $s$ were not, however, so depleted in this region as for MHV. The resolution of the alternative whether this region may function as packaging signal for HCoV-HKU1 or such similarity reflects the evolutionary conservation needs further experimental verification. For HCoV-OC43 the peak in the windows #189–190 was found for $s = 54$. The peak for $s = 24$ was downstream from this region in the windows #193–194 (sites 20737–21060 within the gene coding for 2'-O-methyltransferase).

Except the genome for HCoV-HKU1 containing tandem repeats, the number of pairs with completely coincident TAMGI motifs within the same two other genomes was about 15–20 for each particular step $s$. The pair-wise comparison of coincident motifs in different genomes provides the number of pairs 20–50 for the counterpart $s$ (Supplement S3). Both complete coincidences and coincidences up to one mismatch were mainly significantly higher between HCoV-OC43 and HCoV-HKU1 than between HCoV-OC43 and MHV as well as between HCoV-HKU1 and MHV (Supplement S3). The positioning of the coincident motifs on the genomes was closer for the longer motifs, indicating their conservation during molecular evolution. The longest closely positioned coincident motifs for the pair-wise comparison of the genomes were as follows: OC43-HKU1, $s = 21$, $k = 8$, GAATGAAT (15271, 15531); $k = 9$, TGCTTTTAA (8318, 8578); $s = 57$, $k = 9$, ATAATTATG (14790, 15050); $k = 10$, TTGTTTATAA (14727, 14987); $k = 12$, GGTTGGGATTAT (15149, 15409); $s = 60$, $k = 8$, ATGATTTT (14121, 14381); $s = 66$, $k = 8$, ATAAGTTT (17127, 17387); OC43-MHV, $s = 24$, $k = 8$, TTTTGATG (19819, 20071); $s = 54$, $k = 8$, TAATGTTA (20512, 20761); $k = 9$, TATGTTTTT (16637, 16898); $k = 10$, TATGCTATTA (14939, 15200); $s = 57$, $k = 8$, TGTTTTTA (16639, 16900); $k = 9$, TTGCTTTTA (9760, 10021); $k = 10$, TTTAAGAAGT (16078, 16339); $s = 63$, $k = 13$, GTTTGTTTGTGGA (7729, 7990); $s = 66$, $k = 9$, GCTTTTTGT (9440, 9701); GATGATTTT (20456, 20705); $k = 11$, CATATTATTGT (16305, 16566); HKU1-MHV, $s = 21$, $k = 8$, CTGATGTT (13658, 13659); GTATTAAT (26312, 27194); $s = 24$, $k = 10$, AATGATAATA (17749, 17750); $s = 54$, $k = 8$, TTATGATA (17382, 17383); $k = 9$, ATAAACAAT (16637, 16638); $s = 57$, $k = 8$, GGTTGGGA (15352; 15353); $s = 60$, $k = 12$, TGTTGATGATTT (16092, 16093); $s = 66$, $k = 8$, TGTTGATG (16092, 16093) (see also Supplement

S3). The closely positioned triple coincidences were rather rare and comprised only motifs no longer than $k = 7$. The TAMGI motifs for HCoV-HKU1 also contained transcription regulatory sequences CTAAAC (Woo et al., 2010), $s = 57$, CTAAAC (15155); $s = 66$, ACCTAAAC (26466); this feature was absent for the other two viruses.

*3.4. RNAPS in the genomes of the human and animal alphacoronaviruses*

The human alphacoronaviruses HCoV-229E and HCoV-NL63 cause the common cold in healthy adults (Dijkman & van der Hoek, 2009). Yet close phylogenetically, they shared only 65% sequence identity. Being less virulent than SARS- and MERS-related viruses, these alphacoronaviruses were suggested to use as more safe models for development of drugs against SARS- and MERS-related viruses in laboratory conditions (Chakraborty & Diwan, 2020). The structure of N protein for HCoV-NL63 retains the major structural features inherent to N proteins of the other coronaviruses (Chang et al., 2014; Peng et al., 2020; Szelazek et al., 2017). The mode of oligomerization during formation of nucleocapsid for HCoV-229E is also close to that in the others (Lo et al., 2012).

In this section we study RNAPS in the genomes of HCoV-229E and HCoV-NL63 (accessions: NC_002645.1 and NC_005831.2, respectively) and compare them with RNAPS in the genomes of miniopterus bat coronavirus 1 (Bat-CoV MOP1; accession: EU420138.1), camel isolate (camel α-CoV; accession: KT368906.1), and transmissible gastroenteritis coronavirus in pigs (TGEV; accession: NC_038861.1). If the lengths of the genome and of the gene coding for N protein for SARS-CoV are taken as a reference ($M = 29751$, $L_N = 1269$), their counterparts are distinctly shorter for HCoV-229E, HCoV-NL63, Bat-CoV MOP1, camel α-CoV, and TGEV ($M = 27317$, $L_N = 1170$; $M = 27553$, $L_N = 1134$; $M = 28326$, $L_N = 1170$; $M = 27395$, $L_N = 1149$; and $M = 28586$, $L_N = 1149$, respectively). Therefore, among the other features, we checked a possible shift of the main quasi-periodicity for assembly/packaging motifs from $p = 54$ to $p = 51$.

The related TAMGI deviations within the range of steps $s = 1–550$ and JC spectra for $s_1 = 54$ and $s_2 = 1–600$ are shown in Fig. 9 (on the left and right, respectively; the relevant data for camel α-CoV are presented only in Supplements to save the space). We found that the features for the genome of TGEV were different from those of the other chosen viruses. The difference between the human alphacoronaviruses and TGEV can be attributed to the divergence between respiratory and enteric viruses. Characteristics for TGEV will be described separately below, while the data summary in this passage concerns mainly the first four viruses. The deviation $\kappa_{147}$ was the highest for HCoV-229E, Bat-CoV MOP1, and camel α-CoV, whereas for HCoV-NL63 it was on the ranked position 2 in the same grade as the top rank deviation for $s = 189$ (=21×9), $\kappa_{189} = 10.30$. The step $s = 147$ (and some of the others corresponding to the top ranked deviations) is the multiple of 7×21. The deviations for $s = 21$ were also invariably among top ranked, $\kappa_{21} = 7.72$ (rank position 5 for HCoV-229E); 9.07 (17 for HCoV-NL63); 10.05 (2 for Bat-CoV MOP1); and 8.52 (2 for camel α-CoV). The deviations for $s = 51$ and 54 were distinctly higher than those for $s = 57$, while their mutual ranking depended on the particular

alphacoronavirus, $\kappa_{51}$ = 7.16; 9.10; 9.44; 7.01 and $\kappa_{54}$ = 6.87; 9.31; 7.56; 7.65 for HCoV-229E, HCoV-NL63, Bat-CoV MOP1, and camel α-CoV, respectively. The related deviations for the doubled steps were $\kappa_{102}$ = 4.91; 8.04; 7.81; 5.24 and $\kappa_{108}$ = 6.98; 9.08; 7.74; 6.88, i.e. except Bat-CoV MOP1, were always in favor of quasi-periodicity $p$ = 54. The high deviations for $s$ = 60 appear to be the other common feature for the alphacoronaviruses in this group. The motifs for $s$ = 21, 54, 60 can be considered as fuzzy quasi-periodic ones.

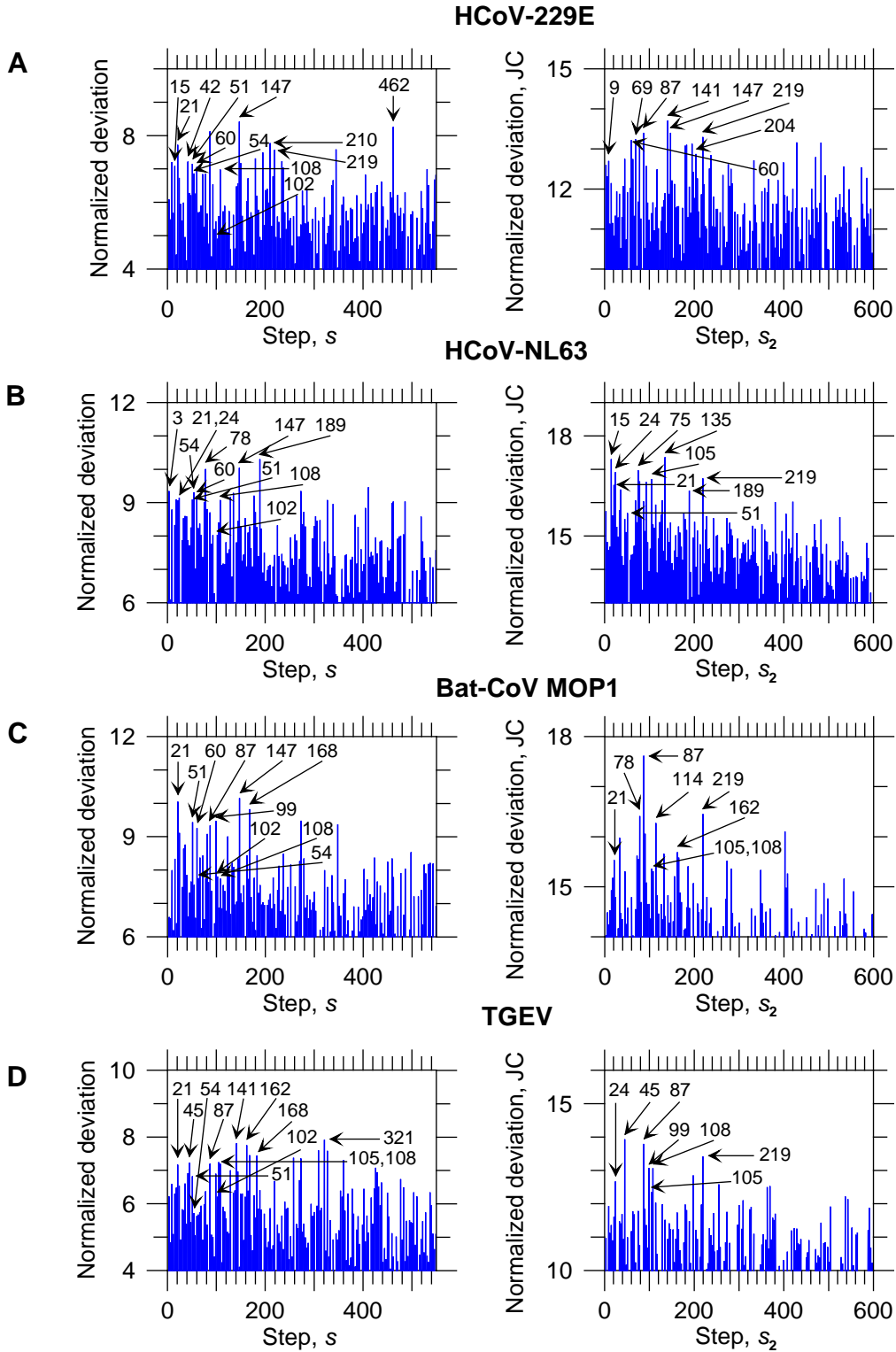

**Figure 9.** The spectra for the TAMGI (on the left) and JC (on the right) normalized deviations defined by Eqs. (7) and (18) for the genomes of the alphacoronaviruses. The TAMGI deviations were calculated within the range $s = 1$–550, whereas the deviations for JC were calculated for $s_1 = 54$ and $s_2 = 1$–600. **A**, HCoV-229E; **B**, HCoV-NL63; **C**, Bat-CoV MOP1; and **D**, TGEV. The characteristic steps $s$ are explicitly marked by arrows.

The JC spectra for $s_1 = 51$ and $s_2 = 1$–600 are shown separately in Fig. 10. The combined TAMGI and JC spectra reveal a trend to the formation of the long 150–450 nt correlational patterns in the genomes of all studied alphacoronaviruses including TGEV. Brief comments on the particular alphacoronaviruses are presented below.

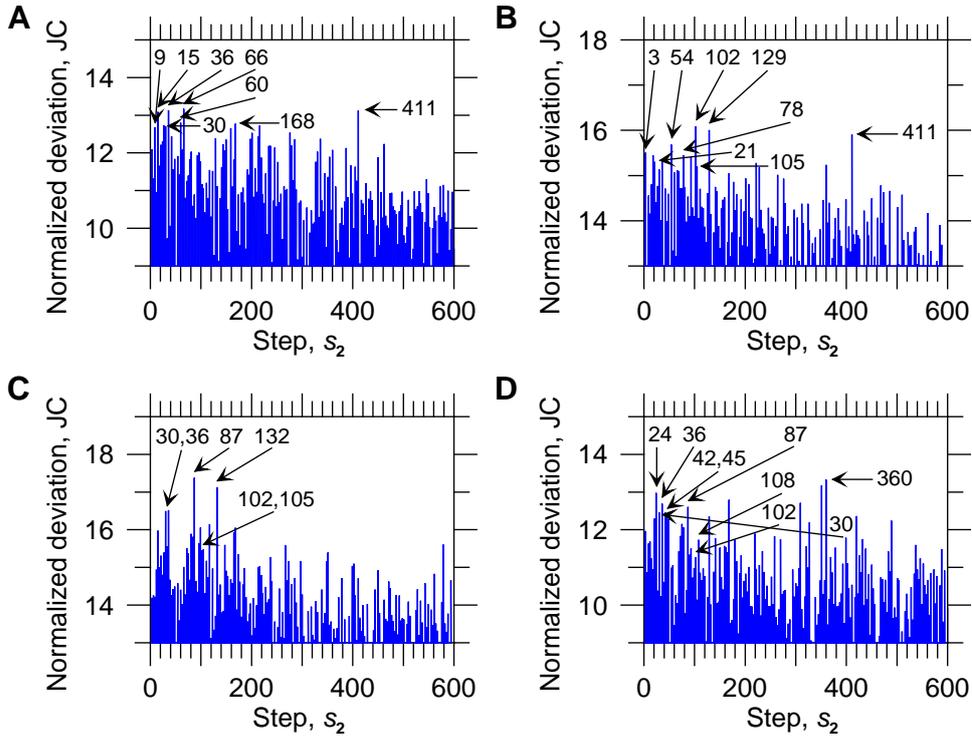

**Figure 10.** The spectra for the JC normalized deviations defined by Eq. (18) for the genomes of HCoV-229E **(A)**; HCoV-NL63 **(B)**; Bat-CoV MOP1 **(C)**; and TGEV **(D)**. The deviations for JC were calculated for $s_2 = 1$–600 and $s_1 = 51$. The characteristic steps $s_2$ are explicitly marked by arrows.

Characteristic steps corresponding to the top ranked deviations for the genome of HCoV-229E are marked by arrows in Figs. 9A and 10A (see also Supplement S1). The RNAPS motifs associated with $s = 54$ were coordinated with the channels 141–87, 87+21, whereas the motifs for $s = 51$ were coordinated with the channels 66–15, 36+15, 60–9, 168–66. Characteristic steps for the genome of HCoV-NL63 are shown in detail in Figs. 9B and 10B. The RNAPS motifs associated with $s = 54$ were coordinated with the channels 78–24, 75–21, 51, 105, whereas the motifs for $s = 51$ were coordinated with the channels 129–78, 54–3, 102, 105. Characteristic steps for the genome of Bat-CoV MOP1 are shown in Figs. 9C and 10C. The RNAPS motifs associated with $s = 54$ were coordinated with the channels 87–33, 87+21, 105, 108, 162 (=54×3), whereas the motifs for $s = 51$ were coordinated with the channels 87–36, 36+15, 105, 102, 132–30.

Characteristic steps for the genome of TGEV are marked by arrows in Figs. 9D and 10D. The top three ranked TAMGI deviations in Fig. 9D, left, corresponded to the steps $s = 321, 141$ (instead of 147 for the other alphacoronaviruses), and 162 (=3×54) and can be attributed to the same grade. The deviation for $s = 51$ was a grade higher in comparison with that for $s = 54$ (6.82 and 5.72, respectively), while both were distinctly higher than the deviation for $s = 57$ ($\kappa_{57} = 5.04$). The related deviations for the doubled steps were $\kappa_{102} = 6.21$ and $\kappa_{108} = 7.20$ (note also $\kappa_{105} = 7.25$ and that $s = 105 = 21\times5 = 51+54$). The rank position for the deviation $\kappa_{21} = 7.17$ was 15 (on the lower boundary of the same grade as for the top rank deviations). The top grade JC deviations $\iota_{21 \wedge s_2}$ corresponded to $s_2 = 141, 162, 183$, and 42 (the latter correlations can be considered as the indicator of quasi-periodicity for $p = 21$), whereas the top grade JC deviations $\iota_{54 \wedge s_2}$ corresponded to $s_2 = 45, 87, 219, 99$, and 108 (the latter correlations can be considered as the indicator of quasi-periodicity for $p = 54$). Thus, the motifs for $s = 21$ and 54 can be considered as fuzzy quasi-periodic patterns. The RNAPS motifs associated with $s = 54$ were coordinated with the channels 99–45, 108, 105, and 87+21, whereas the motifs associated with $s = 51$ were coordinated with the channels 30+21, 87–36, 360–309, and 24+30.

The profiles for the characteristic TAMGI motifs in the genomes of HCoV-229E, HCoV-NL63, Bat-CoV MOP1, and TGEV are shown in Fig. 11. We begin with the latter profiles because there is an experimentally verified packaging signal for TGEV (Escors et al., 2003; Morales et al., 2013). As proved by Morales et al. (2013), the first 598 nt from the 5'-end of the genome function as the packaging signal. The last 494 nt at 3'-end enhance the packaging efficiency yet are not crucial for the packaging. As is seen from Fig. 11D, the most pronounced peak in the packaging signal region was detected for the motifs $s = 51$ associated with the fuzzy helical periodicity in the genomic sequences. The abundance of these motifs downstream of the packaging signal (see windows #6, 10–14) facilitates the encapsidation. The motifs for $s = 51$ may act synergistically with the motifs for $s = 24, 141, 162$, and 321. The motifs with $k \geq 6$ in the region of packaging signal were as follows (see also Supplement S3): $s = 21$, AGTAGG (start at site 198); $s = 24$, CAAGATC (332), GATCAC (486), TGATCTT (524); $s = 51$, TGGCTAT (27), TTTGTC (106), TTCTAT (113), GGACAAGCG (130), TTCGTAC (441); $s = 54$, TTTTAAA (3), TCCGCC (249), CCGCCAG (304); $s = 60$, TCTTCTT (539); $s = 141$, AGGACA (129), CGCCTAGTA (226), TGCCTAGT (367), GATCAC (486), CAATGG (593); $s = 162$, AACAATTC (325), GTTTTCGT (438); $s = 321$, ATTCAG (151), TTGGAA (172), TTGAGG (427), CATTGG (497). The motifs for $s = 60, 141, 162$, and 321 were abundant in the region near 3'-end. Among other features, note high peaks within region of windows #30–32 coding for nsp3.

The abundance of motifs near the 5'-end was detected for $s = 21, 24$, and 147 (HCoV-229E; Fig. 11A); $s = 60$ and 147 (HCoV-NL63; Fig. 11B); $s = 21, 54$, and 147 (Bat-CoV MOP1; Fig. 11C), whereas the abundance of motifs near the 3'-end was detected for $s = 24$ (HCoV-229E; Fig. 11A); $s = 54$ (HCoV-NL63; Fig. 11B); $s = 21, 54$, and 60 (Bat-CoV MOP1; Fig. 11C). By analogy with TGEV, only the profiles for Bat-CoV MOP1 may be associated with the packaging signal in the region near the 5'-end. Generally, the motifs in 5'- and 3'-end regions may be associated with other functional mechanisms (Yang & Leibowitz, 2015). Note also high peaks on the profiles for $s = 24$ and 54 in the region of the gene coding for N protein in the genome of HCoV-NL63 (windows #245–246 and #248–249, respectively, in Fig. 11B) and for $s = 51$ in the regions of the genes coding for HCNV63gp2 and HCNV63gp3 proteins (or S protein and protein 3, respectively; windows #212–213 and #231–232). A similar high peak in the region of the gene coding for S protein was detected for the profile Bat-CoV MOP1, $s = 51$ (windows #230–231 in Fig. 11C).

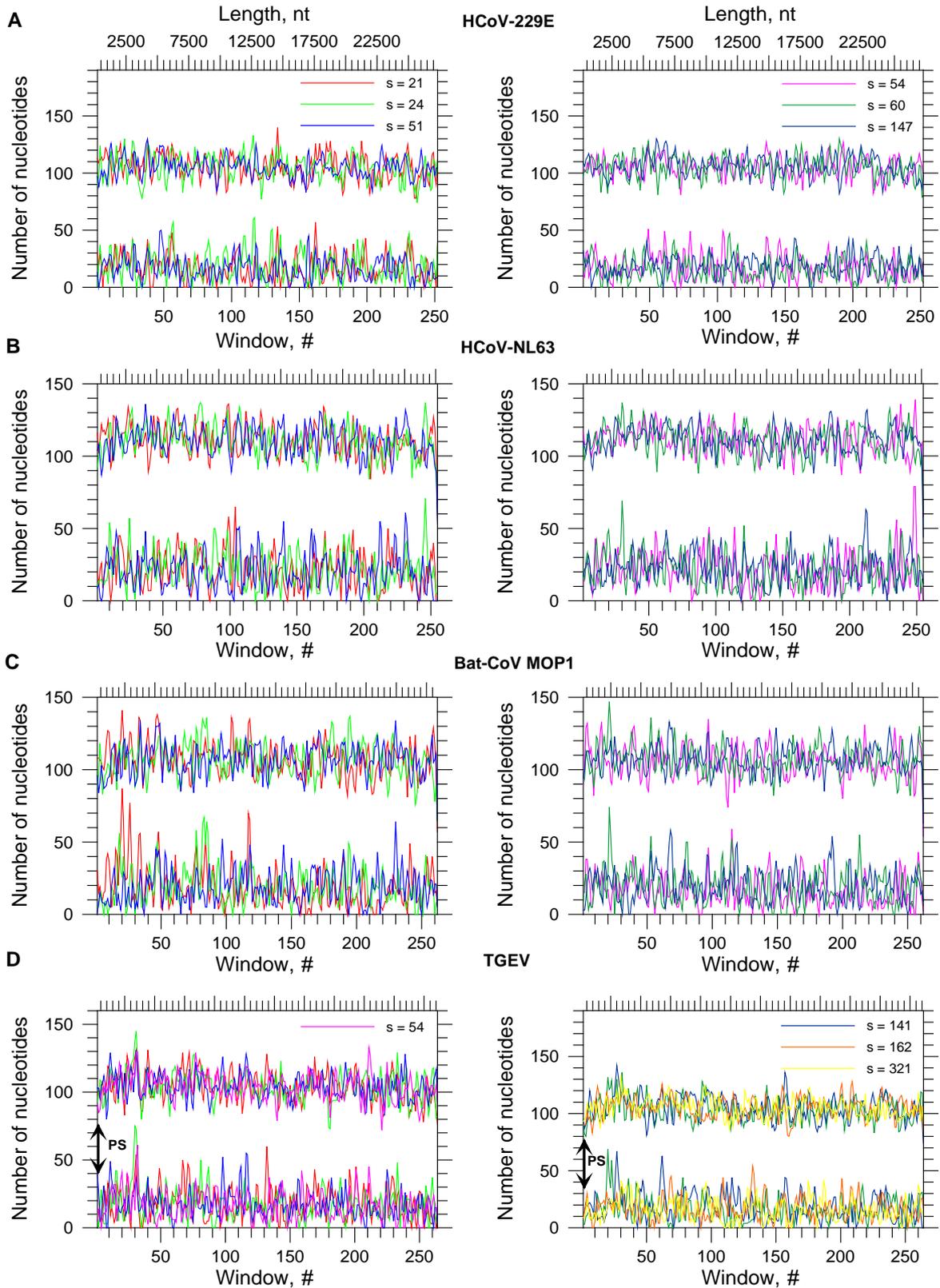

**Figure 11.** The distributions of motifs over the genomes of HCoV-229E **(A)**; HCoV-NL63 **(B)**; Bat-CoV MOP1 **(C)**; and TGEV **(D)** calculated with windows of width $w = 216$ and sliding step of 108 (half-overlapping windows). The corresponding TAMGI steps for the profiles were as follows; on the left, $s = 21, 24, 51$ (all four genomes), while for TGEV $s = 54$ was added; on the right, $s = 54, 60, 147$ (for the genomes of HCoV-229E, HCoV-NL63, Bat-CoV MOP1), while for TGEV the profiles corresponded to the steps $s = 60, 141, 162, 321$. The upper profiles correspond to all motifs ($k \geq 1$), whereas the lower profiles correspond to $k \geq 5$. The two-sided arrows for TGEV **(D)** indicate the middle of the experimentally verified packaging signal located at the sites 1–598.

The motifs with $k \geq 6$ are summarized in Supplement S3. Some motifs reveal remarkable conservation on both the characters and positioning on the genome. In particular, the following triple correspondences for HCoV-229E, HCoV-NL63 and camel α-CoV were detected: $s = 54$, $k = 9$, CAGATCCTA (with the start positions 14975, 14894, 14990, respectively); $s = 60$, $k = 9$, GTGTTTGTT (8121, 8037, 8136); $s = 147$, $k = 8$, AAACTGGT (9945, 9864, 9960) as well as many correspondences for the shorter motifs. Such triple correspondences for HCoV-229E, HCoV-NL63 and Bat-CoV MOP1 were absent. Similar correspondences were found only for the pair-wise comparison, e.g., for HCoV-229E and Bat-CoV MOP1, $s = 24$, $k = 8$, GTGGTTCA (9594, 10143); $s = 60$, $k = 9$, ATGGAAATG (19674, 20175) or for HCoV-NL63 and Bat-CoV MOP1, $s = 21$, $k = 8$, TTGTAATG (7054, 7672); $s = 51$, $k = 10$, TTGCTAATGA (15560, 16169); $s = 60$, $k = 8$, TGTTGATG (14928, 15537); $s = 147$, $k = 10$, TTTATGAATT (20108, 20702).

We studied also coincidences of 6-mer motifs for the counterpart steps in the different genomes (Supplement S3). We restricted ourselves to the complete coincidences and to the coincidences up to one mismatch. In the latter case the complete coincidences were discarded. The complete coincidences included also the self-coincidences (i.e. each motif was counted at least once). Each pair of coincident motifs with different positions on the genome was counted once. It was found within such a scheme that more than a half of motifs for HCoV-229E and camel α-CoV were completely coincident. The pair-wise complete coincidences for the other alphacoronaviruses were at the level 10–20%. The complete coincidences were commonly the lowest between motifs for TGEV and the other four viruses, with the minimum of coincidences for the steps $s = 51$ and 54 associated with the packaging periodicity. The complete coincidences for the motifs in the genome of HCoV-NL63 were the highest with Bat-CoV MOP1 and were at the level about 20%.

## 4. Discussion

### *4.1. Comparison of RNAPS in the genomes of different coronaviruses*

TAMGI motifs are intrinsically related to the general organization of the genome, RNAPS being one of the elements of the genome organization. The choice of putative targets in viral genomes is strongly hampered by high frequency of point mutations and indels. As has been proved earlier, TAMGI motifs are robust with respect to point mutations and indels (Chechetkin & Lobzin, 2020, 2021). Let the length of the genome be *M*. If the step of TAMGI is *s* and the number of indels is $N_{ID}$, they affect the correlation motifs in the region $2sN_{ID}$ around indels. The fraction of modified motifs is $2sN_{ID}/M$. This imposes a restriction $s < M/2N_{ID}$ for the approximate conservation of primary motifs in the presence of indels. For the steps *s* < 500–600 used in this paper, the maximum number of indels should not exceed 25–30, whereas the real numbers of indels are about 1–5. Therefore, the longer motifs with $k \geq 5–6$ can be used as putative therapeutic targets. It would be insightful to compare the impact of mutations related to the longer motifs with $k \geq 5–6$ on virus viability as the impact from mutations within reconstructed motifs and/or from mutations leading to the elongation of motifs is expected to be stronger.

The overlapping TAMGI motifs generate the complicated network of coordinated motifs. The RNAPS around the step $s = 54$ can be affected by the other overlapping motifs for the different steps and vice versa. The character of overlapping can be assessed by the Jaccard coefficients. Combining the normalized deviations for TAMGI and JC provides results directly in terms of underlying motifs within genomic sequences. The vocabulary of motifs and their positioning in the genome can be conveniently assessed against the annotated genomic sequences and specialized data bases (Ibrahim et al., 2018; Sharma et al., 2016).

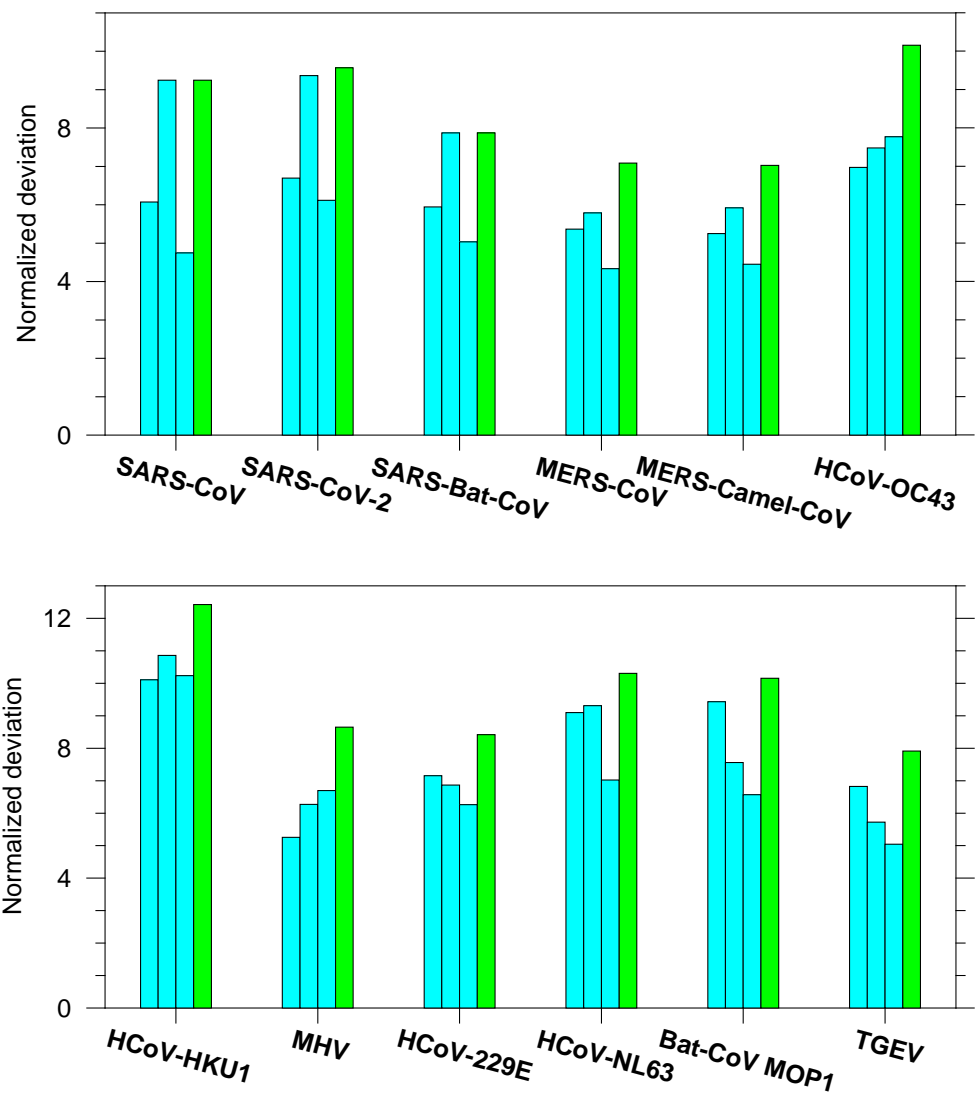

**Figure 12.** The normalized TAMGI deviations associated with RNAPS in the genomes of the different coronaviruses in comparison with the maximum deviation. The first three columns correspond to the normalized

TAMGI deviations for $s$ = 51, 54, and 57, respectively, whereas the extreme right column corresponds to the maximum deviation.

The TAMGI deviations for the steps $s$ = 51, 54 and 57 in the genomes of different coronaviruses are mutually compared in Fig. 12. These data reveal the approximate bias to $s$ = 51 and to $s$ = 57 for the genomes shorter and longer than that of SARS-CoV. The deviations for $s$ = 54 were the highest throughout related TAMGI spectra for the lineage B betacoronaviruses comprising the most virulent coronaviruses SARS-CoV, SARS-CoV-2, and SARS-Bat-CoV, whereas for the other lineages the relevant deviations associated with RNAPS can be a grade or two lower in comparison with the maximum one (see Fig. 12). Though N proteins are evolutionary conserved on the coding and structural levels, the mechanisms of evolving RNAPS are distinctly more variable for the different coronaviruses.

The comparison of motif profiles for the different steps $s$ with the experimentally verified packaging signals for MERS-CoV, MHV, and TGEV showed that commonly the motifs of one type are abundant within the region of the packaging signal, whereas the other motifs are depleted. The type of the putative indicator motifs and their peaked location in the genome vary strongly for the particular viruses. This can be attributed to the molecular mechanisms of the friend-or-foe recognition which should be highly specific. A part of peaks and troughs in the motif profiles can be attributed to the mechanisms of initiation and elongation of encapsidation, replication and gene expression.

The spectra for TAMGI and JC deviations and related profiles for motifs depend on the particular lineage and on the particular virus within lineage. However, some of the features appear to be highly reproducible for all coronaviruses. As has been discussed in Section 3.3, the region corresponding to the windows #29–32 (sites 3025–3564) is occupied by 30-nt tandem repeats in the genome of the coronavirus HCoV-HKU1. This region is within the gene coding for nsp3 for all coronaviruses. The long stretches with tandem repeats are simultaneously a source of a variety of correlation TAMGI motifs which can participate in the different molecular mechanisms (see Fig. 8B, Supplement S3 and general discussion by Chechetkin & Lobzin (2021)). Though the region with such tandem repeats is inherent only to HCoV-HKU1, the peaks in the corresponding regions were also found for the profiles of motifs throughout all coronaviruses (see Figs. 2, 5, 8, and 11). These peaks were pronounced in the motif profiles for $s$ = 54 associated with the period of helical nucleocapsid for the most pathogenic coronaviruses SARS-CoV and SARS-CoV-2, whereas for MERS-CoV the pronounced peak in this region was for $s$ = 105. Targeting of the quasi-repeating motifs in both genomic RNA or/and related proteins should be more efficient because of the larger statistical weight of such motifs that makes nsp3 a promising therapeutic target.

### *4.2. Abundance of N proteins*

The complete encapsidation of the SARS-CoV genome needs $4.4 \times 10^3$ N proteins if the nucleocapsid helix pitch is associated with 54 nt (Chechetkin & Lobzin, 2020). This result agrees with cryo-EM data (Chang et al., 2014; Gui et al., 2017). The fuzzy shifts of periodicity $p$ = 54 to 57 for the longer genomes of the lineage A betacoronaviruses and to 51 for the shorter genomes of alphacoronaviruses retain the estimate above. This means that the expression of N gene should be approximately the same for all coronaviruses. Our bioinformatic analysis and cryo-EM data indicate that N proteins should be the most abundant. The abundant N proteins induce strong immune response and the antibodies against N proteins can be used for the diagnostics of coronaviral infection and for the development of vaccines (Chang et al., 2016; Dutta et al., 2020).

There are also indirect indications in favor of such conclusion based on the mechanism of expression of structural and accessory proteins in coronaviruses via nested subgenomic mRNAs (sgmRNAs) (Sawicki et al., 2007; Wu & Brian, 2010; Yang & Leibowitz, 2015; Zhang et al., 1994). The order of location of the genes coding for the structural proteins on the genome is the same for all coronaviruses and corresponds to S-E-M-N. The nested sgmRNAs serve for the expression of the genes S-E-M-N, E-M-N, M-N, and N. The conclusion on the abundance of N proteins contradicts to the estimates obtained via stoichiometry (Neuman & Buchmeier, 2016).

The specific N-RNA interactions are retained after freeing viral RNA from nucleocapsid. The biomolecular N-RNA condensates provide a plausible model for N-RNA complexes in cellular cytosol (Cascarina & Ross, 2020; Iserman et al., 2020; Lu et al., 2021; Perdikari et al., 2020; Savastano et al., 2020). A part of TAMGI motifs associated with the helical ribonucleocapsid periodicity $p = 54$ and with the other pronounced correlations should be responsible for the specific N-RNA interactions in such biomolecular condensates.

The specific N-RNA interactions after freeing viral RNA from nucleocapsid proved also to affect the replication and transcription (see for a review Chang et al., 2014; Masters, 2019; and references therein). The level of gene expression should be concordant with the number of protein copies needed for self-assembly of the membrane envelope with the spike glycoproteins and the packaging of the genome. This implies some regulation feedback mechanisms for expression. The experimental data and TAMGI motif profiles indicate that N proteins can mediate the regulation.

*4.3. Recombination and encapsidation*

Together with frequent mutations, the recombination is one of the major source of virus variability and diversity. The recombination region for coronaviruses is mainly attributed to the receptor-binding domain of the spike (S) glycoprotein (Bobay et al., 2020; Tao et al., 2017; V'kovski et al., 2021; Zhu et al., 2020). Taking into account the order of the structural genes in the genome, S-E-M-N, the fragment downstream from such recombinant region should code for all structural and accessory proteins. In this section we discuss compatibility of recombination and encapsidation. Consider a recombinant genome combined by ORF1ab coding for RNA replicase (fragment I) and sgRNA coding for all structural and accessory proteins (fragment II). The experimentally established packaging signals in the genomes of coronaviruses are located within the fragment I (Escors et al., 2003; Fosmire et al., 1992; Hsin et al., 2018; Kuo & Masters 2016; Morales et al., 2013), whereas fragment II codes for M and N proteins though homologous yet different in some features from their counterparts in the primary genome without recombination. All experiments and the results of our bioinformatic analysis indicate that the specific recognition between the genomic RNA and M/N proteins is needed for the proper transport of RNA into the membrane envelope and encapsidation. In the case of recombinant genome such a recognition should be between RNA from the fragment I and M/N proteins translated from the fragment II. It should be expected that the efficient RNA transport, encapsidation and turnover of the whole virus life cycle need that homology between the fragments II of the virus genome and/or the similarity between structures of M/N proteins before and after recombination ought to exceed some threshold needed to make fuzzy the friend-or-foe recognition between the RNA and M/N proteins after recombination. In some cases the 3'-end may also affect the efficiency of packaging (Morales et al., 2013; Yang & Leibowitz, 2015). This means that a recombinant virus should be suboptimal (if viable) from the packaging point of view in the overwhelming majority of cases. The recombinant virus can be adopted and selectively optimized during subsequent evolution or be extinct (Jensen & Lynch, 2020; Vignuzzi & López, 2019). A part of the pronounced peaks for the motif profiles in the region of the gene coding for S protein may be related to recombination. As is seen from Figs. 2, 5, 8, and 11, this feature is typical of the coronavirus genomes.

The location of the packaging signal within the recombinant fragment II coding for the structural and accessory proteins would make more compatible recombination and encapsidation. It is interesting to investigate whether the peaked region 28729–29052 within the gene coding for N protein in the genome of SARS-CoV-2 (see the profile for $s = 9$ in Fig. 2B; windows #267–268) affects the packaging. The other peaked region 13933–14256 (windows #130–131 in Fig. 2B) for $s = 9$ lies within the gene coding for nsp12. The similar peaked regions 23005–23328 (windows #214–215 in Fig. 2A) and 6589–6912 (windows #62–63 in Fig. 2A) for $s = 9$ in the genome of SARS-CoV were found within the genes coding for S protein and for nsp3, respectively. The neighboring peaks for $s = 54$ and 9 in the windows # 113 and 114 (sites 12097–12420 on the profiles shown in Figs. 2A and 2B) were reproducible for both SARS-CoV and SARS-CoV-2 and lie within the gene coding for nsp8. The synergistic action of these regions on packaging cannot be excluded as well. Note the depletion of the other motifs in all noted peaked regions for $s = 9$. The corresponding motifs for $s = 9$, $k \geq 6$ can be found in Supplement S3.

*4.4. Co-infection and encapsidation*

The recombination needs co-infection of one host cell by two or more coronaviruses. Another aspect of co-infection is related to competitive interactions between M/N proteins and genomic RNAs of two or more coronaviruses within one host cell (Sungsuwan et al., 2020). The molecular mechanisms of co-infection for human coronaviruses are still poorly understood (Chaung et al., 2020; Lai et al., 2020; Ou et al., 2020). The specific friend-or-foe recognition between RNA and M/N proteins should play important role during co-infection as well.

*4.5. Drug-targeting RNAPS and N proteins and their use for vaccines*

A year after outbreak of COVID-19 pandemic, the efforts of scientists leaded to the development of several vaccines ready or nearly ready for practical medical applications (Artese et al., 2020; Dong et al., 2020; Karpiński et al., 2021; Li & Li, 2021; Logunov et al., 2021; Padron-Regalado, 2020; Singh & Gupta, 2020). In this section we discuss several possibilities related to RNAPS. As has been established in Section 3.1, the transcription regulatory sequences with the motifs TACGAACTT coordinated with the helical capsid periodicity $p = 54$ were strictly conserved for SARS-CoV, SARS-related Bat-CoV and SARS-CoV-2 and were similarly located with respect to the starts of the genes coding for E proteins. Targeting of this motif by any agent (ligands, proteins, aptamers, modified N proteins or their fragments) should switch off the expression of E gene and interrupt the virus life cycles simultaneously for all three viruses. The same applies to the motif CCCTTTGTC (27596) nearby the start of the gene coding for E protein in the genome of MERS-CoV (Section 3.2). The replacements of the transcription regulatory sequences (rewiring regulatory network) proved to attenuate the virulence of SARS-CoV (Graham et al., 2018). The total number of relatively long motifs with $k \geq 9$ does not exceed 15–30 for the particular steps (Supplement S3). The targeting of such long motifs may appear to be also efficient for attenuation of viral activity. Note in particular, the long 17-mer motif TATTCAAACAATTGTTG (start at 3268) within the gene coding for nsp3 or 12-mer motif ACTGCCACTAAA (29060) within the gene coding for N protein (both for $s = 54$) in the genome of SARS-CoV-2. Hybridizing the RNA palindrome CAATTG within 17-mer motif with targeted DNA can be used for cutting this site by restrictase.

Unlike progress in the development of vaccines, the development of antiviral drugs remains still much less successful (Artese et al., 2020; Rohilla, 2021; Twomey et al., 2020). Search for agents efficiently targeting RNAPS or N proteins would be highly desirable. Other strategies can be based on preventing oligomerization of N proteins (He et al., 2004; Chang et al., 2013, 2016). N proteins strongly binding with RNA can affect various regulatory mechanisms with the RNA participation. In particular, it has been shown that SARS-CoV N protein can activate an AP-1 pathway, which regulates many cellular processes, including cell proliferation, differentiation, and apoptosis (He et al., 2003; Zhu et al., 2021). It was also shown that binding of SARS-CoV-2 N proteins to the host 14-3-3 protein in the cytoplasm can regulate nucleocytoplasmic N shuttling (Tugaeva et al., 2021). The multifunctional role of N proteins can be important not only for the virus life cycle but for the pathways and general disease progression as well.

*4.6. Virus transmission and motifs*

The risks of the animal-to-human and human-to-animal transmission of viruses can be assessed not only by the general homology between related genomic sequences for the human and animal coronaviruses but also by the correspondence between functional motifs. We found that some of the motifs were highly reproducible both in characters and positioning on the genome throughout the lineages (Supplement S3). Among studied genomes, the closest homology was between the camel and human MERS-CoV. They were nearly coincident and differed as one isolate differs from another. In this case it can be said about adaptation of the animal virus within the human organism. Two other pairs with the closest correspondence between motifs were human and bat SARS-CoV as well as HCoV-229E and camel α-CoV. The homology between the camel and human coronaviruses for MERS-CoV and HCoV-229E enhances the risk of co-infection by these viruses.

## 5. Conclusion

Combining detection and reconstruction of correlational and quasi-periodic motifs by TAMGI provides a convenient tool for the study of viral genomic sequences. Supplementing TAMGI by the Jaccard coefficients elucidates the relationships within intricate network of mutually connected regulatory motifs. In the case of coronaviruses such unified technique displays the relationships between RNAPS and the other motifs within genome organization. The first, more general, level of characterization of viral genomes comprises the ranked TAMGI deviations and the study of mutual relationships between them by JC. These characteristics remained mainly reproducible within the same lineage of the coronaviruses but differed between the lineages. The second, more specific, level of characterization is related to the repertoires of longer TAMGI motifs (typically with $k \geq 5$–$6$), their sequential ordering and distribution over the genome. In this form, the TAMGI-JC technique can be applied to the assessment of the evolutionary divergence between viruses and to the problem of subtyping. The application of the TAMGI motifs to the subtyping of viruses is close to the general discriminant genomic analysis with $k$-mers (Ounit et al., 2015; Tomović et al., 2006). Motifs reconstructed by TAMGI can also be used for the assessment of the mutation impact and for therapeutic targeting. Further progress in the study of (multi)functional role of TAMGI motifs can be achieved by the additional experimental work. The vocabulary of functional motifs for the coronaviruses is of basic interest and can help in the practical medical applications.